\documentstyle[12pt]{article}

\def\qe{=}

\def\bigans{y }
\ifx\answ\bigans\input epsf.tex
\else\fi

\def\scrip{{\cal I}^+} \def\scrim{{\cal I}^-}
\def\sigp{\sigma^+} \def\sigm{\sigma^-} \def\yp{y^+}
\def\ym{y^-} \def\be{\begin{equation}}
\def\ee{\end{equation}} \def\bal{\begin{array}{l}}
\def\eal{\end{array}} \def\bea{\begin{eqnarray}}
\def\eea{\end{eqnarray}}  \def\om{\omega}
\def\lam{\lambda} \def\lxp{\lambda x^+} \def\xp{x^+}
\def\lxm{\lambda x^-} \def\xm{x^-} \def\lxbp{\lambda
\bar{x}^+} \def\lxbm{\lambda \bar{x}^-} \def\al{\alpha}
\def\alb{\bar{\alpha}} \def\ml{\frac{M}{\lambda}}
\def\mbl{\frac{\overline{M}}{\lambda}} 
 \def\half{\frac{1}{2}}
\def\bra{\langle} \def\ket{\rangle}
\def\sigbp{\bar{\sigma}^+} \def\ybm{\bar{y}^-}
\def\epsilon{{\Delta M}} \def\0{P} \def\q{1} \def\h{g}
\def\bra{\langle} \def\m{{\cal M}} \def\mb{{\bar{\cal M}}}
\def\ket{\rangle} \def\lra{} \def\m{{\cal M}}
\def\mb{{\bar{\cal M}}} \def\bb{\bar b} \def\l{\lambda}
\def\xp{x^+} \def\xm{x^-} \def\xbp{\bar \xp}  \def\r{r}\def\f{\bar f}\def\g{f}

\begin{document}
\begin{titlepage}

\begin{flushright}
MIT-CTP-2341\\
hep-th/9408039\\
July 1994
\end{flushright}

\vskip 0.9truecm

\begin{center}
{\large {\bf Breakdown of the Semi-Classical Approximation \\
at the Black Hole Horizon$^*$}}
\end{center}

\vskip 1.5cm

\begin{center}
{Esko Keski-Vakkuri, Gilad Lifschytz, Samir D. Mathur and Miguel E.
Ortiz$^\dagger$}
\vskip 0.4cm
{\it Center for Theoretical Physics,
    Laboratory for Nuclear Science \\
    and Department of Physics \\
    Massachusetts Institute of Technology \\
    Cambridge MA 02139, USA.\\
    e-mail: keskivakkuri, gil1, mathur or ortiz @mitlns.mit.edu }
\end{center}

\vskip 0.7cm

\begin{center}
Submitted to: {\it Physical Review D}
\end{center}

\vskip 1.3cm

\noindent {\small {\bf Abstract:}
The definition of matter
states on spacelike hypersurfaces of a 1+1 dimensional black
hole spacetime is considered. The effect of small quantum
fluctuations of the mass of the black hole due to the
quantum nature of the infalling matter is taken into
account. It is then shown that the usual approximation of
treating the gravitational field as a classical background
on which matter is quantized, breaks down near the black
hole horizon. Specifically, on any hypersurface that
captures both infalling matter near the horizon and Hawking
radiation, quantum fluctuations in the background geometry
become important, and a semiclassical calculation is
inconsistent. An estimate of the size of correlations
between the matter and gravity states shows that they are so
strong that a fluctuation in the black hole mass of order
$e^{-M/M_{Planck}}$ produces a macroscopic change in the
matter state.
}

\rm
\noindent
\vskip 2.5cm

\begin{flushleft}
$^*$ {\small This work was supported in part by funds
provided by the U.S. Department of Energy (D.O.E.) under
cooperative agreement DE-FC02-94ER40818.}\\ $^\dagger$
{\small Address from September 1994: Dept. of Physics, Tufts
University, Medford MA 02155, USA}
\end{flushleft}

\end{titlepage}

\section{Introduction}

Since the original papers of Hawking \cite{Haw,Haw2} arguing
that black holes should radiate thermally, and that this
leads to an apparent loss of information, it has been hoped
that investigations of this apparent paradox would lead to a
better understanding of quantum gravity. Over the last few
years, there has been renewed interest in this general
problem.  One reason is the construction of 1+1 dimensional
models where evaporating black holes can  be easily studied
\cite{cghs}. Another reason is the work by 't Hooft
\cite{hooft,hooft2,shw} suggesting that the black hole
evaporation process may not be semiclassical. This idea is
based in part on the fact that although Hawking radiation
emerges at low frequencies of order $M^{-1}$ at
$\cal{I}^{+}$, it originates in very high frequency vacuum
modes at $\cal{I}^{-}$ and even close to the black hole
horizon, the latter frequencies being about $e^M$ times the
Planck frequency \cite{jacobson} (here $M$ is the mass of
the black hole in Planck units). 't Hooft also argues that
if the black hole evaporation process is to be described by
unitary evolution, then there should exist large commutators
between operators describing infalling matter near the
horizon and those describing outgoing Hawking radiation
\cite{shw} despite the fact that they may be spacelike
separated.

Recently\footnote{Historically, the greatest champions of
this view point have been Page \cite{page} and 't Hooft
\cite{hooft,hooft2}.} Susskind {\it et. al.} have argued
that the information contained in infalling matter could be
transferred to the Hawking radiation at the black hole
horizon, thus avoiding information loss \cite{stu}. A common
argument against this possibility is that from the
perspective of an infalling observer, who probably sees
nothing special at the horizon, there is no mechanism that
could account for such a transfer of information. In
response, Susskind suggests a breakdown of Lorentz symmetry
at large boosts, and a principle of {\em complementarity}
which says that one can make observations either far above
the horizon or near the horizon, but somehow it should make
no sense to talk of both \cite{stu,suss2}.

The 1+1 dimensional black hole problem including the effects
of quantum gravity was recently studied in Ref. \cite{verl}.
It was found that there are very large commutators between
operators at the horizon, and  operators at $\cal{I}^{+}$
measuring the Hawking radiation, agreeing with the earlier
work of 't Hooft \cite{shw}. Ref. \cite{verl} assumes a
reflection boundary condition at a strong coupling boundary.
Some natural modifications of this boundary condition have
been studied recently in \cite{dmst}. There have been many
other studies of quantum gravity on the black hole problem,
some of which are listed in \cite{qgbh}.

Let us recall the basic structure of the black hole problem
\cite{Haw2}. Collapsing matter forms a black hole, which
then evaporates by emission of Hawking radiation \cite{Haw}.
The radiation carries away the energy, leaving `information'
without energy trapped inside the black hole. The Hawking
radiation arises from the production of particle pairs, one
member of the pair falling into the horizon and the other
member escaping to form the Hawking radiation outside the
black hole. The quantum state of the quantum particles
outside the black hole is thus not a pure state, and one may
compute the entanglement entropy between the particles that
fall into the black hole and the particles that escape to
infinity. It is possible to carry out such a computation
explicitly in the simple 1+1 dimensional models referred to
above. One finds \cite{KM,fpst} that this entropy equals the
quantity expected on the basis of purely thermodynamic
arguments \cite{zurek}.

Such calculations are carried out in the semiclassical
approximation, where one assumes that the spacetime is a
given 1+1 dimensional manifold, and the matter is given by
quantum fields propagating on this manifold. How accurate is
this description? We wish to examine the viewpoint raised by
't Hooft and Susskind (referred to above) that quantum
gravity is important in some sense at the horizon of the
black hole. To this end we start with a theory of quantum
gravity plus matter, and see how one obtains the
semiclassical approximation where gravity is classical but
matter is quantum mechanical. The extraction of a
semiclassical spacetime from suitable solutions of the
Wheeler--DeWitt equation has been studied in
\cite{semiclass}. Essentially, one wishes to obtain an
approximation where the variables characterizing gravity are
`fast' (i.e. the action varies rapidly with change of these
variables) and the matter variables are `slow' (i.e. the
action varies slowly when they change). This separation
hinges on the fact that the gravity action is multiplied by
an extra power of the Planck mass squared, compared to the
matter variables, and this is a large factor whenever the
matter densities are small in comparison to Planck density.
We recall that the matter density is indeed low at the
horizon of a large black hole (this is just the energy in
the Hawking radiation). One might therefore expect the
semiclassical approximation to be good at the horizon. It is
interesting that this will turn out {\em not} to be the
case, as we shall now show in a 1+1 dimensional model.

It was suggested in \cite{mathur} that a semiclassical
description ({\em i.e.} where gravity is classical but
radiating matter quantum) can break down after sufficient
particle production. This suggestion is based on the fact
that particle creation creates decoherence \cite{cm}, but on
the other hand an excess of decoherence conflicts with the
correlations between position and momentum variables needed
for the classical variable \cite{pazsinha}. In this paper we
investigate this crude proposal and find that there is
indeed a sense in which the semiclassical approximation
breaks down near a black hole. It turns out that the
presence of the horizon is crucial to this phenomenon, so
what we observe here is really a property of black holes.

Since in black hole physics one is interested in concepts
like entropy, information, and unitarity of states, it is
appropriate to use a language where one deals with `states'
or `wavefunctionals' on spacelike hypersurfaces, instead of
considering functional integrals or correlation functions
over a coordinate region of spacetime. In this description,
the dynamical degrees of freedom are 1-geometries, and it is
more fundamental to speak of the state of matter on a
1-geometry than on an entire spacetime. Thus, we will need
to study the canonical formulation of 1+1 dimensional
dilaton gravity. Recall that in this theory the gravity
sector contains both the metric and an additional scalar
field, the dilaton, which together define a 1-geometry. The
space of all possible 1-geometries is called superspace. We
assume that our theory of quantum gravity plus matter is
described by some form of Wheeler--DeWitt equation
\cite{wdw}, which enforces the Hamiltonian constraint on
wavefunctionals in superspace. For dilaton gravity alone, a
point of superspace is given by the fields
$\{\rho(x),\phi(x)\}$. Here we have assumed the notation
that the metric along the 1-dimensional geometry is
$ds^2\qe e^{2\rho} dx^2$, and $\phi$ is the dilaton. One of the
constraints on the wavefunctionals  is the diffeomorphism
invariance in the coordinate $x$. Using this invariance we
may reduce the description of superspace so  that different
points just consist of intrinsically different 1-geometries.
More precisely, choose any value of $\phi$, say $\phi_0$.
Let $s$ denote the proper distance along the 1-geometry
measured from the point where $\phi\qe \phi_0$, with $s$
positive in the direction where $\phi$ decreases. The
function $\phi(s)$ along the 1-geometry describes the
intrinsic structure of the 1-geometry, and is invariant
under spatial diffeomorphisms (we have assumed here for
simplicity that $\phi$ is a monotonic function along the
1-geometry, and that the value $\phi_0$ appears at some
point along the 1-geometry). Loosely speaking, we may regard
superspace as the space of all such functions $\phi(s)$ (for
a spacetime with boundary, this description must be
supplemented with an embedding condition at the boundary).

Let us now consider the presence of a massless scalar field
$f(x)$. Points of superspace now are described by
$\{\phi(s),f(\phi(s))\}$, and wavefunctionals on this space,
$\Psi[\phi(s),f(\phi(s))]$, satisfy the Wheeler--DeWitt
equation
\be
(H_{\rm gravity}~+~H_{\rm matter})\Psi[\phi(s),f(\phi)]~\qe ~0
\ .
\label{intro1}
\ee
We are now faced with the question: How do we obtain the
semiclassical limit of quantum gravity, starting from some
theory of quantum gravity plus matter? At the present point
we have only 1-geometries in the description, and we have to
examine how the 1+1 dimensional spacetime emerges in some
approximation from $\Psi[\phi(s),f(\phi)]$. Obtaining a 1+1
dimensional spacetime has been called the `problem of time'
in quantum gravity, and considerable work has been done on
the semiclassical approximation of gravity as a solution to
this problem \cite{semiclass}.  We wish to reopen this
discussion in the context of black hole physics.

In mathematical terms, we have $\Psi[\phi(s),f(\phi(s))]$
giving the complete description of matter plus gravity. What
is the state of matter on a time-slice? If we are given a
classical 1+1 spacetime, then a time-slice is given by an
intrinsic 1-geometry $\phi(s)$ (plus a boundary condition at
infinity). Thus the matter wavefunctional on a time-slice
$\phi(s)$ should be given by
\be
\Psi_{\phi(s)}[f(\phi(s))]~\equiv~\Psi[\phi(s),f(\phi(s))] \ ,
\label{intro2}
\ee
The semiclassical approximation then consists of
approximating the full solution of the Wheeler--DeWitt
equation by the product of a semiclassical functional of the
gravitational variables alone, times a matter part which is
taken to be a solution
\be
\psi^{\m}_{\phi(s)}[f(\phi(s))]
\label{intext1}
\ee
of the functional Schr\"odinger equation on some mean
spacetime $\m$ (here the function $\phi(s)$ is like a
generalized time coordinate on $\m$).  If any quantum field
theory on curved spacetime calculation using
(\ref{intext1}) can be used to approximate the result
obtained using the exact solution of the Wheeler-DeWitt
equation of (\ref{intro2}), then we say that the
semiclassical approximation is good. On the other hand, if
this approximation fails to work, we conclude that quantum
fluctuations in geometry are important to whichever question
it is that we wished to answer.

For the black hole problem, it is appropriate to make a
separation between the matter regarded as forming the black
hole, denoted by $F(\phi(s))$, and all other matter
$f(\phi(s))$. It is then more natural to regard $F(\phi(s))$
as part of the gravitational degrees of freedom, and it is
certainly regarded as a classical  background field in the
derivation of Hawking radiation using the semiclassical
approximation. In this situation we must be more precise
about what we require for the semiclassical approximation to
be good. Assume that the black hole is formed by the
collapse of some wavepacket of matter $F$, into a region
smaller than the Schwarzschild radius. We note that the
energy of this matter wavepacket cannot be exactly $M$,
because an eigenstate of energy would not evolve at all over
time in the manner needed to describe the collapsing packet.
In fact, since the matter will be localized to within  the
Schwarzschild radius $M$, there will be a momentum
uncertainty much greater than $1/M$ in Planck units, which
leads to an energy uncertainty which must also be much
larger than $1/M$. This uncertainty is still quite small,
but should nevertheless not be ignored. The different
possible energy values in this range $(M,M+\epsilon)$ where
$\epsilon \gg 1/M$, will give different semiclassical
spacetimes. For the semiclassical approximation to be good
for any given computation, it must be independent of which
of the slightly different spacetimes is chosen. Conversely,
if the difference in any quantity of interest becomes
significant when evaluated on different spacetimes in the
above mass range, then we cannot use a mean 2-geometry to
describe physics, and we should say that the semiclassical
approximation is not good\footnote{The role of fluctuations
in the mass of the infalling matter was also discussed in
\cite{verl}. Generally, fluctuations in geometry can also
arise from other sources, but we shall ignore these here.}

Casting this problem in the language of the preceeding
paragraphs, we must ask whether the wavefunctional of matter
from the full quantum solution of the Wheeler--DeWitt
equation is well approximated by working on a fixed
spacetime $\m$ of mass $M$ and ignoring the uncertainty
$\epsilon$ in $M$. Now, suppose that the semiclassical
approximation were a good one when describing the state of
matter on a given time-slice $\phi(s)$. If we consider the
different matter states that are obtained on $\phi(s)$ by
taking different values for $M$, which cannot be clearly
distinguished because we are averaging over the fluctuations
in geometry, then these states should not be `too different'
if there is to be an unambiguous definition of the state on
the time-slice. This is a minimal requirement for a
semiclassical calculation to be a good approximation to
$\Psi[\phi(s),F(\phi(s)),f(\phi(s))]$.

Let the state of quantized matter obtained by working on
$\m$ be $\psi^\m_{\phi(s)}[f(\phi(s))]$,  where in $\m$ the
energy of the infalling matter is $M$. This is a state in
the Schr\"{o}dinger representation, and thus depends on the
time-slice specified by the function $\phi(s)$ (plus
boundary condition). At slices corresponding to early times
(i.e. near $\scrim$, before the black hole formed) for all
spacetimes with mass $M$ in the range $(M,M+\epsilon)$, we
fix the matter state to be approximately the same in each
spacetime. In terms of a natural inner product relating
states on a common 1-geometry in different spacetimes (which
we define in this paper), this means that
\be
\langle\psi_{\phi(s)}^{\m}|\psi_{\phi(s)}^{\mb}\rangle
 \approx 1
\label{intro3}
\ee
on these early time slices, where $\mb$ is a spacetime with
mass $\bar M$ in the above range.  On each spacetime the
matter state evolves in the Schr\"{o}dinger picture in
different ways,  so that the inner product (\ref{intro3})
will not be the same on all slices. For the semiclassical
approximation to be good at any given slice, we need that
(\ref{intro3}) hold on that slice.

\ifx\answ\bigans
\medskip
\begin{center}
\leavevmode
\epsfxsize 2.5in
\epsfbox{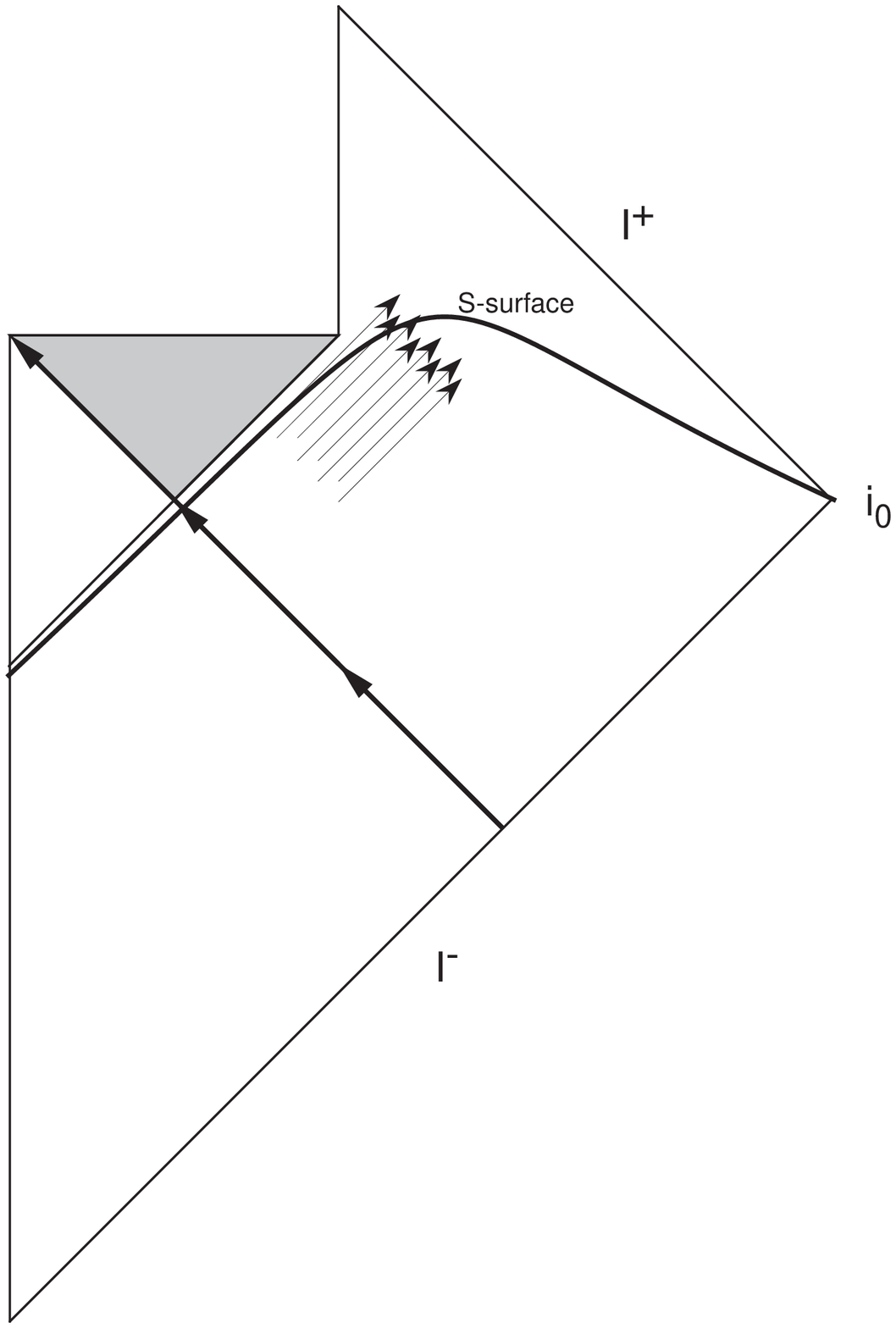}
\end{center}
\vskip 0.5 true cm
\begin{quotation}
\small
\noindent
{\bf Figure 1:}
An example of an S-surface, shown in an evaporating black hole
spacetime.
\end{quotation}
\medskip
\else
\fi

Having fixed the matter states on different spacetimes so
that they are very similar at early times, we  analyse later
time slices to check that this property still holds. Any
slice is taken to start at some fixed base point near
spatial infinity. Consider now a slice that moves up in time
near $\scrip$ to capture some fraction of the Hawking
radiation. The slice then comes to the vicinity of the
horizon, and then moves close to the horizon, so as to reach
early advanced times before entering the strong coupling
domain (see Fig. 1). The importance of such slices to the
black hole paradox has been emphasized by Preskill
\cite{preskill} and Susskind {\it et. al.} \cite{stu} in
their arguments relating to information bleaching and to the
principle of black hole complementarity. Susskind {\it et.
al.} conjectured that the large Lorentz boost between the
two portions of the slice should lead to a problem in the
semiclassical description of a black hole. Slices of this
type have also been used in the literature as part of a
complete spacelike slicing of spacetime, that stays outside
the horizon of the black hole \cite{lines} and captures the
Hawking radiation, and on which semiclassical physics should
therefore apply. For these surfaces, which we shall refer to
as S-surfaces, we shall show in this paper that it is no
longer the case that matter states are approximately the
same for different background spacetimes. Indeed, even for
$|M-\bar M|\sim e^{-M}$ we find that on a 1-geometry
$\phi(s)$ of this type,
$$
\langle\psi_{\phi(s)}^{\m}|\psi_{\phi(s)}^{\mb}\rangle
 \approx 0.
$$
As was argued above, the fluctuations in the mass of the
hole must be at least of  order $\Delta M >1/M$, so we see
that the state of matter on such slices is very ill defined
because of the fluctuations in geometry. This shows that at
least one natural quantity that we wish to consider in black
hole physics, the state of matter on what we have termed an
S-surface, is not given well by the semiclassical
approximation.

The plan of this paper is the following. In section 2 we
review the CGHS model, and give some relevant scales. In
section 3 we study the embedding of 1-geometries in
different 1+1 dimensional semiclassical spacetimes. In
section 4 we compare states of matter on the same
1-geometry, but in different spacetimes. Section 5 is a
general discussion of the meaning of these results and of
possible connections to other work.

\section{A review of the CGHS model}

There follows a quick review of the CGHS model \cite{cghs},
with reference to the RST model \cite{rst} which includes
back-reaction and defines some relevant scales in the CGHS
solution. Although all calculations in this paper are for a
CGHS black hole, the general features of the results that
are derived are expected to apply equally well to other
black hole models in two and four dimensions.

The Lagrangian for two dimensional string-inspired dilaton
gravity is
\be
S_G\qe {1\over 2\pi}\int
dxdt\;\sqrt{-g}\;e^{-2\phi}\left[R+4(\nabla\phi)^2+4\lambda^2\right]
\label{1}
\ee
where $\phi(x)$ is the dilaton field and $\lambda$ is a
parameter analogous to the Planck scale. Writing
$$
ds^2\qe -e^{2\rho}dx^+ dx^-
$$
where $x^{\pm} \qe t\pm x$ are referred to as Kruskal
coordinates, (\ref{1}) has static black hole solutions
\be
e^{-2\rho}\qe e^{-2\phi}\qe {M\over\lambda}-\l^2\xp\xm
\label{2}
\ee
and a linear dilaton vacuum (LDV) solution with $M\qe 0$. More
interesting is the solution obtained when (\ref{1}) is
coupled to conformal matter,
$$
S\qe S_G-{1\over 4\pi}\int dxdt\;\sqrt{-g}\;(\nabla f)^2,
$$
where $f$ is a massless scalar field. A left moving shock
wave in $f$ giving rise to a stress tensor
$$
{1\over 2}\partial_+f\partial_+f\qe M\delta(\xp-1/\l)
$$
yields a solution
\be
e^{-2\rho}\qe e^{-2\phi}\qe -{M\over\lambda}(\l\xp-1)
\Theta(\xp-1/\l)-\l^2\xp\xm
\label{3}
\ee
representing the formation of a black hole of mass $M/\l$ in
Planck units (the Penrose diagram for this solution in shown
in Fig. 2). For $\l\xp<1$ (region I), the solution is simply
the LDV, whereas the solution for $\l\xp>1$ (region II),
$$
e^{-2\rho}\qe e^{-2\phi}\qe {M\over\lambda}-\l\xp\left(\l\xm+{M\over\l}
\right)
$$
is a black hole with an event horizon at $\l\xm\qe -M/\l$.

\ifx\answ\bigans
\medskip
\begin{center}
\leavevmode
\epsfxsize 3in
\epsfbox{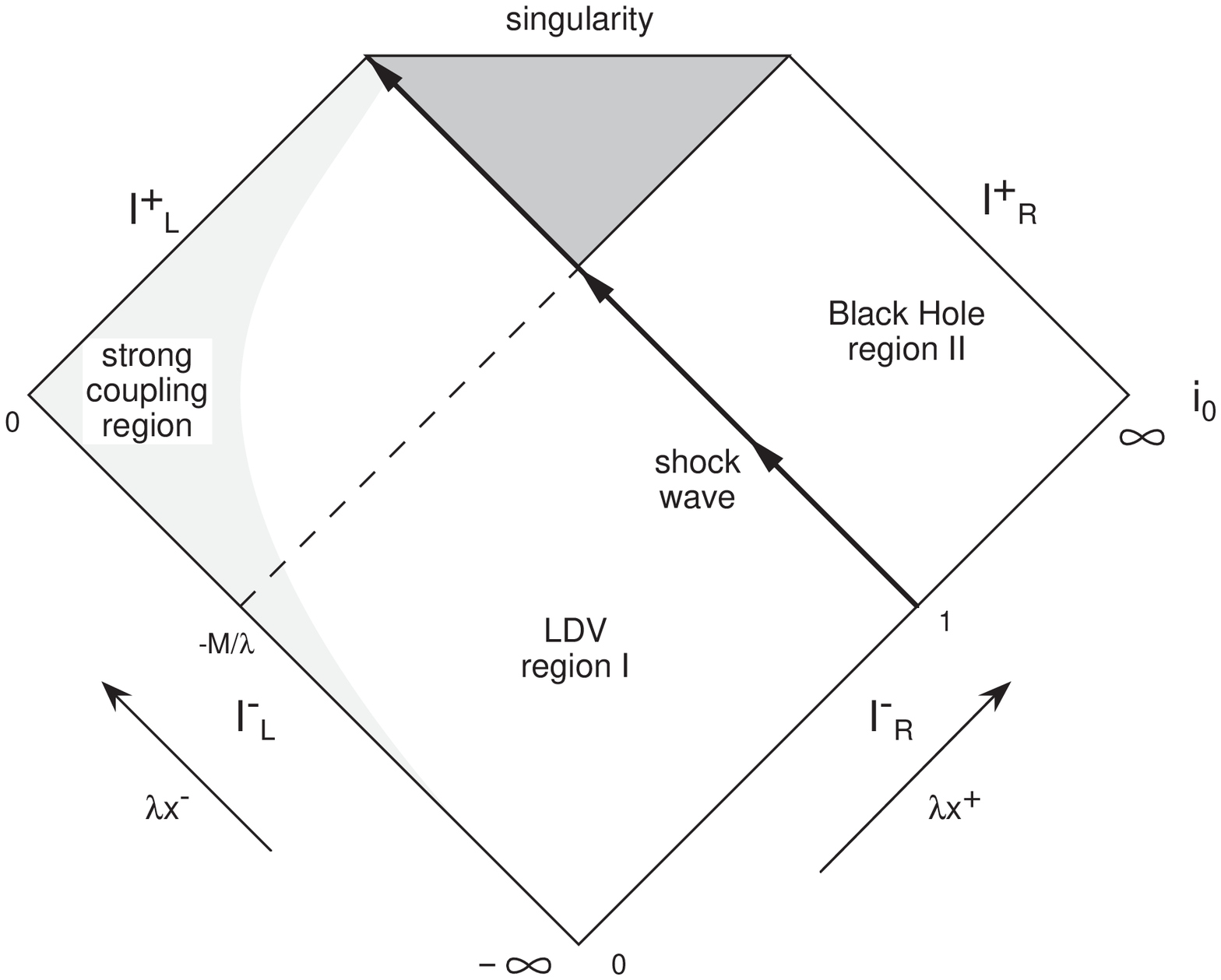}
\end{center}
\vskip 0.5 true cm
\begin{quotation}
\noindent
\small {\bf Figure 2:} The Penrose diagram of the CGHS solution.
\end{quotation}
\medskip
\else
\fi

It is possible to define asymptotically flat coordinates in
both regions I and II. In region I, we define
\be
\l\xp\qe e^{\l y^+},\qquad\l\xm\qe -\ml e^{-\l y^-}
\label{coord1}
\ee
and in region II we introduce the ``tortoise'' coordinates
$\l \sigma^{\pm}$:
\be
\l\xp\qe
e^{\l \sigma^+},\qquad\l\xm+{M\over \l}\qe -e^{-\l \sigma^-} \
\ .
\label{coord2}
\ee
The coordinate $y^-$ is used to define right moving modes at
$\scrim_L$. To define left moving modes at $\scrim_R$ we can
use either $\yp$ or $\sigp$. As (\ref{coord1}) and
(\ref{coord2}) tell us, both coordinates can be extended to
$I\cup II$ so that $\yp \qe  \sigp$. It is easy to see that as
$\sigma\to\infty$ or as $y\to\infty$, $\rho\to -\infty $.
Notice also that $e^{\phi}$ plays the role of the
gravitational coupling constant in this theory. It is
generally believed that semiclassical theory is reliable in
regions where this quantity is small. At infinity $e^\phi\to
0$, and so this is a region of very weak coupling. Even at
the horizon, $e^\phi\qe \sqrt{\l/M}$ is small provided that the
mass of the black hole is large in Planck units ($M/\l\gg
1$). This is assumed to be the case in all calculations so
that the weak coupling region extends well inside the black
hole horizon.

One virtue of this two dimensional model is that it is
straightforward to include the effects of backreaction by
adding counterterms to the action $S$. This was first done
by CGHS, but a more tractable model was introduced by RST
who found an analytic solution for the metric of an
evaporating black hole. However, the RST model still
exhibits all the usual paradoxes associated with black hole
evaporation (for a review see \cite{preskill,gidd}).

\ifx\answ\bigans
\medskip
\begin{center}
\leavevmode
\epsfxsize 2.5in
\epsfbox{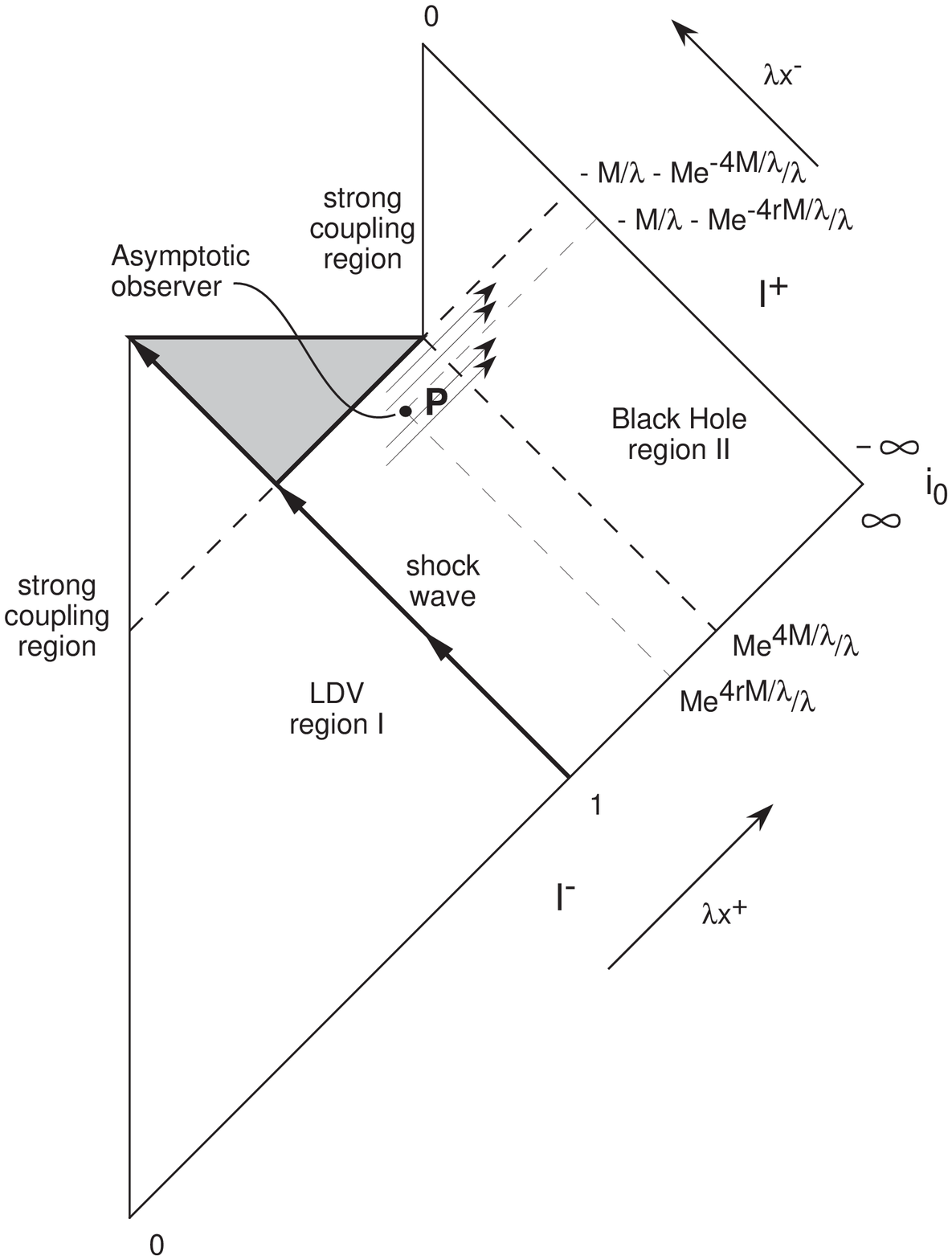}
\end{center}
\vskip 0.5 true cm
\begin{quotation}
\noindent
\small {\bf Figure 3:} The Penrose diagram of the
RST solution with some
approximate scales shown.
\end{quotation}
\medskip
\else
\fi

Although we will carry out our calculations in the simpler
CGHS model, the RST solution (whose Penrose diagram is shown
in Fig. 3), is a useful guide for identifying certain scales
in the evaporation process. These can be usefully carried
over to a study of the CGHS solution, and serve to determine
the portion of that solution that is unaffected by
backreaction: The time scale of evaporation of the hole as
measured by an asymptotic observer is $t_E\sim 4M$ in Planck
units; the value of $\xm$ at which a proportion $r$ of the
total Hawking radiation reaches ${\cal I}^+$ is
$\l\xm_P\qe -M(1+e^{-4r M/\l})/\l$ (by this we mean that the
Hawking radiation to the right of this value carries energy
$r M$); the value of $\xp$, for $\xm\qe \xm_P$, which
corresponds to a point well outside the hole, in the sense
that the curvature is weak and the components of the stress
tensor are small is $\l\xp_P\qe Me^{4rM/\l}/\l$ provided that
$\l\xp>e^{2M/\l}$. On the basis of these scales, we can
define a point $P$ at $(\l\xp_P,\l\xm_P)$ as defined above,
located just outside the black hole, in the asymptotically
flat region, and to the left of a proportion $r$ of the
Hawking radiation.

\section{Embedding of 1-geometries}

In this section, we shall compare how a certain spacelike
hypersurface $\Sigma$ may be embedded in collapsing black
hole spacetimes (\ref{3}) of masses $M$ (denoted by $\m$)
and $\bar M\qe M+\Delta M$ (denoted by $\mb$), where $\Delta M$
is a fluctuation of at most Planck size.

In 1+1 dimensional dilaton gravity models an invariant
definition of a 1-geometry is provided by the value of the
dilaton field $\phi(s)$ as a function of the proper distance
$s$ along the 1-geometry, measured from some fixed reference
point. For   spacetimes with boundary, such as the black
hole geometries in the CGHS model, this reference point may
be replaced by information about how the  1-geometry is
embedded at infinity. It is natural to regard asymptotic
infinity as a region   where hypersurfaces can be nailed
down by external observers who are not a   part of the
quantum system we are considering. We impose the   condition
that 1-geometries in different spacetimes should   be
indistinguishable for these asymptotic observers, ensuring
that the semiclassical approximation holds for these
observers. This   condition and the function $\phi(s)$ are
enough to define a unique map of $\Sigma$ from ${\cal M}$ to
$\bar{\cal M}$.

It is important to point out at this stage that it is
possible that this map is not  well defined for some $\mb$,
in the sense that there may exist no spacelike hypersurface
in $\mb$ with the required properties. For the surfaces we
consider, this issue does not arise. Further, it can be
argued that there is no important effect of this phenomenon
on the state of the matter fields, at least as long as one
is away from strong curvature regions. (To see this it is
helpful to use the explicit quantum gravity wavefunction for
dilaton gravity given in \cite{kunst}). For this reason we
shall ignore all spacetimes $\mb$ where $\Sigma$ does not
fit.

Given an equation for $\Sigma_\m$,
$$
\lambda x^-\qe f(\l x^+)
$$
and expressions for $\rho(x^+,x^-)$ and $\phi(x^+,x^-)$ in
${\cal M}$ and $\bar\rho(\bar{x}^+,\bar x^-)$ and
$\bar\phi(\bar x^+,\bar x^-)$ in $\bar{\cal M}$, we
determine the corresponding equation for $\Sigma_{\bar {\cal
M}}$,
$$
\lambda \bar x^-\qe \bar f(\l \bar x^+)
$$
by requiring that $\phi(s)\qe \bar\phi(\bar s)$ and similarly
$d\phi/ds(s)\qe  d\bar\phi/d\bar s(\bar s)$ (it is if these
equations have no real solution for a given $\mb$ that we
say that $\Sigma$ does not fit in $\mb$). These conditions
require one boundary condition which fixes $\Sigma_\mb$ at
infinity, and this may be chosen in such a way that the
equations for $\Sigma_\m$ and $\Sigma_\mb$ are the same in
asymptotically flat (tortoise) coordinates sufficiently far
from the black hole.

We shall demonstrate that while most surfaces embed in very
slightly different ways in spacetimes $\m$ and $\mb$ with
masses differing only at the Planck scale, there is a
special class of surfaces for which this is not true (what
we mean by embeddings being different will be discussed
later). These are the S-surfaces which catch both the
Hawking radiation (the Hawking pairs reaching ${\cal I}^+$,
but not those ending up at the singularity) and the
in-falling matter near the horizon (see Fig. 1). It is
useful to give an example of such surfaces. A straight line
in Kruskal coordinates $x^\pm$ going through a point
$P\sim(Me^{4rM/\l}/\l,-M(1+e^{-4rM/\l}))$, is a line of this
type, catching a proportion $r$ of the outgoing Hawking
radiation, provided the slope of the line is extremely small
-- of order $e^{-8r M/\l}$. The smallness of this parameter
will play an important role in our discussion.  Although the
line is straight in Kruskal coordinates, it will, of course,
look bent in the Penrose diagram, ending up at $i_0$. Far
from the horizon, these lines are lines of constant
Schwarzschild time $\l t\qe 4\r M$, giving an interpretation
for minus one half the logarithm of the slope in terms of
the time at infinity.

It is worth pointing out that the map from a surface
$\Sigma_\m$ in $\m$ to the corresponding surface
$\Sigma_\mb$ in $\mb$ defines a map from any point $Q$ on
$\Sigma_\m$ to a point $\bar Q_\Sigma$ on $\Sigma_\mb$ in
$\mb$. Any other choice of surface $\Xi_\m$ in $\m$ passing
through $Q$ maps $Q$ to a different point ${\bar Q}_\Xi$ in
$\mb$. This uncertainty in the location of a point $\bar Q$
in $\mb$ gives a geometric way of defining the fluctuations
in geometry around $Q$. Generally, we may expect all the
images of $Q$ in $\mb$ to lie within a small region of
Planck size. However, we shall see below that this is not
the case near a black hole horizon.

\subsection{Basic Equations}

Here we present the basic equations describing the embedding
of $\Sigma$. In a collapsing black hole manifold $\m$ of
mass $M$ (\ref{3}), it is convenient to define $\Sigma$ as
$$
\lambda x^{-} \qe \g( \lambda x^{+} ) -M/\lambda.
$$
If we use Kruskal coordinates $\bar x^\pm$ to describe
$\Sigma$ in a black hole manifold $\mb$ of mass $\bar M$ as
$$
\lambda \bar{x}^{-} \qe \f ( \lambda \bar{x}^{+} ) -\bar{M}/\lambda,
$$
then in region II of (\ref{3})
\begin{eqnarray}
&& \frac{M}{\lambda} - \lambda x^{+} \g (\lambda x^{+}) \qe
 \frac{\bar{M}}{\lambda} - \lambda \bar{x}^{+} \f (\lambda
 \bar{x}^{+})
\label{g1}
\\
&& \frac{\g + \lambda x^{+} \g' }{\sqrt{-\g'}} \qe
\frac{\f + \lambda \bar{x}^{+} \f' }{\sqrt{-\f'}}
\label{g2}
\end{eqnarray}
where prime denotes a derivative with respect to the
argument. The first equation is the requirement of equal
$\phi(s)$ and the second of equal ${d \phi }/{ds}(s) $.

Once we identify the embedding of $\Sigma$ in $\mb$, we can
then identify points in both spacetimes by the value of $ s
$ on $\Sigma$. This identification may be described by the
function $ \bar{x}^{+}( x^{+} ) $ between coordinates on
$\Sigma$ in each of the spacetimes. To solve the equations
(\ref{g1}) and (\ref{g2}), for $ \bar{x}^{+}( x^{+} ) $,
differentiate (\ref{g1}) by $ x^{+} $  and divide by
(\ref{g2}), to get
$$
\frac{d \bar{x}^{+}}{dx^{+}} \qe \frac{\sqrt{-\g'}}{\sqrt{-\f'}} .
$$
Another combination of these equations gives
$$
\sqrt{-\f'} \qe \frac{-(\g + \lambda x^{+}\g')\pm
\sqrt{(\g-\lambda x^{+}
\g')^{2}-4
\epsilon \g'/\l}}{2\lambda \bar{x}^{+} \sqrt{-\g'}}
$$
where $\epsilon \qe \bar{M}-M$. Combining both equations,
\be
\ln (\lambda \bar{x}^{+} )\qe
2 \int d (\lambda x^{+} ) \frac{\g'}{(\g + \lambda x^{+} \g') \mp
\sqrt{(\g-\lambda x^{+}\g')^{2} - 4 \epsilon \g'/\l}}
\label{g3}
\ee
which is a general expression for $\bar x^+(x^+)$ for any
$\Sigma$. Similarly, if we label the one geometry by $
\lambda x^{+} \qe \h ( z^{-} ) $ where  $ z^{-}\qe \lambda x^{-}
+M/\lambda $ (using the notation $ \h\qe \g^{-1} $), we find an
analogous expression for $ \bar{x}^{-} ( x^{-} ) $:
\be
\ln (\lambda \bar{x}^{-}+ \bar{M}/\lambda )\qe 2 \int d z^{-}
\frac{\h'}{(\h +  z^{-} \h') \mp
\sqrt{(\h- z^{-} \h')^{2} - 4 \epsilon \h'/\l}}  \ \ ,
\label{g3a}
\ee
In (\ref{g3}) and (\ref{g3a}), the sign of the square root
is determined by requiring that as $ \epsilon $ tends to
zero we get $ \bar{x}^{\pm}\qe x^{\pm} $. From these equations
one can construct the corresponding one geometry in $\mb$.
In order for the solution to make sense, the expressions
inside the square root must be positive. This condition is a
manifestation of the fitting problem mentioned  above.

\subsection{A large shift for straight lines}

For simplicity, we focus our attention on lines that are
straight in the Kruskal coordinates $x^\pm$. Below we
present a quick analysis of the embedding of these
1-geometries in neighbouring spacetimes. In the next
subsection a more detailed treatment will be given.

Consider the line $\Sigma$ defined in $\m$ by the equation
$$
\lambda x^{-}\qe
\g(\l x^+)-{M\over\l}\qe -\alpha^{2} \lambda x^{+} +b.
$$
It is easy to see that as a consequence of (\ref{g1}) and
(\ref{g2}), the function $\f(\l \bar x^+)$ describing the
deformed line in Kruskal coordinates on $\mb$ must also be
linear. This is a helpful simplification. Let us write the
equation for $\Sigma$ in $\mb$ as
$$
\lambda \bar{x}^{-}\qe \f(\l\bar x^+)-{\bar M\over\l}\qe
-\bar{\alpha}^{2} \lambda\bar{x}^{+} +\bar b
$$
The parameters $b$ and $\bb$ are related by
\be
\frac{(\bar{b}+\bar{M}/\lambda)^2}{\bar{\alpha^{2}}}\qe
{4\epsilon\over\l} +{(b+M/\lambda)^{2}
\over\alpha^{2} }
\label{g4}
\ee
It is useful to define another quantity $\delta$, so that
$\Sigma$ crosses the shock wave, ($\l x^+\qe 1$) in $\m$ at
$\l\xm\qe -M/\l-\delta$ ({\it i.e.} $\delta\qe\alpha^2-b-M/\l$).
We then find from equation (\ref{g3}) that
$$
2\bar{\alpha}\l\bar{x}^{+}\qe 2\alpha \l x^{+} +\delta/\alpha-\alpha \pm
\sqrt{(\alpha^{2}-\delta)^{2}/\alpha^{2} +4\epsilon/\l}
$$

We still have a free parameter $ \bar{\alpha} $. The way to
fix it is by imposing the condition that $\Sigma$ should be
the same for an asymptotic observer at infinity, meaning
that as expressed in tortoise coordinates $\sigma$ or
$\bar{\sigma}$, $\Sigma$ should have the same functional
form up to unobservable (Planck scale) perturbations. This
may be achieved, as we will see later, simply by picking a
point on $\Sigma$ in $\m$, call it $x^{+}_{0}$, and
demanding that both lines have the same value of $ \phi $ at
the point $ x^{+}\qe\bar{x}^{+}\qe x^{+}_{0}$. Then
\be
\bar{\alpha}\qe\alpha  + \frac{\delta/\alpha-\alpha \pm
\sqrt{(\alpha^{2}-\delta)^{2}/\alpha^{2} +4\epsilon/\l}}{2x^{+}_{0}}
\label{g5}
\ee
Taking $x_0\to \infty$ fixes the line at infinity. The
result does not depend on whether we take $x_0\to\infty$ or
just take it to be in the asymptotic region $x_0>M
e^{2M/\l}/\l$.

We can actually derive some quite general conclusions about
how the embedding of $\Sigma$ changes from $\m$ to $\mb$
from (\ref{g4}) and (\ref{g5}). Let us split the possible
$\Sigma$'s into three simple cases, for any value of
$\alpha$ and  $\delta $ (recall that $|\epsilon/\l |<1$):
\begin{enumerate} \item $ (\alpha^{2}-\delta)^{2}/\alpha^{2}
\gg 4\epsilon/\l $

In this case
$$
\bar{\alpha}\qe \alpha
$$
and
$$
\l \bar{x}^{+} \qe \l x^{+}+\frac{\epsilon}{
\l(\alpha^{2}-\delta)} \ \ .
$$

\item $ (\alpha^{2}-\delta)^{2}/\alpha^{2} \ll 4\epsilon/\l $

For $ \epsilon/\l \geq 0 $ (this is taken to avoid fitting
problems)
$$
\bar{\alpha}\qe \alpha
$$
and
$$
\l \bar{x}^{+} \qe \l x^{+} \pm {\sqrt{\epsilon/\l}\over
\alpha } \ \ .
$$

\item $ (\alpha^{2}-\delta)^{2}/\alpha^{2} \sim 4\epsilon/\l $

Again $ \epsilon/\l \geq 0 $ , and we find a similar result
$$
\alpha\qe\bar{\alpha}
$$
and
$$
\l \bar{x}^{+} \sim \l x^{+} \pm{\sqrt{\epsilon/\l}\over\alpha}
\ \ .
$$
\end{enumerate}
In the last two cases the sign $ \pm $ depends on the sign
of $ \alpha^{2}- \delta $.

The above results all show that the slope $\bar\alpha$ of
the line in $\mb$ is virtually identical to the slope
$\alpha$ in $\m$ (identical in the limit $x_0\to\infty$). It
is also the case that the position of the line in the $\xm$
direction is almost the same in $\m$ and $\mb$. However, for
lines with small values of $\alpha$ and $\delta$, there is a
large shift in the location of the line in the
$x^+$ direction in $\mb$ relative to its position in
$\m$. The lines for which this effect occurs are precisely
the S-surfaces that we have discussed above. These were
defined to have $\alpha^{2}\sim Me^{-8r M/\lambda}/\lambda$,
and $ 0 \leq \delta \leq Me^{-4r M/\lambda}/\lambda $, which
are both small enough to compensate for the $\epsilon$ in
the numerator in the expressions above. The large shift, and
the fact that it occurs only for a very specific class of
lines, precisely the S-surfaces which capture both a
reasonable proportion of the Hawking radiation and the
infalling matter (see Fig. 1), is the fundamental result
behind the arguments presented in this paper. The fact that
only a special class of lines exhibit this effect is
reassuring, as it means that any effects that are a
consequence of this shift can only be present close to the
black hole horizon.

\subsection{Complete hypersurfaces}

So far we have not taken the hypersurfaces to be complete,
{\em i.e.}, we have not done the full calculation of
continuing them to the LDV and finishing at infinity in the
strong coupling regime. We will now perform the full
calculation for a certain class of hypersurfaces. They will
provide us a convenient example (for calculational purposes)
for use in section 4, where we will discuss the implications
of the large shift on the time evolution of matter states.

We choose, for convenience, to work with a class of
hypersurfaces that all have $d\phi/ds\qe -\l$:
\be
      \lxm \qe \left\{ \bal \displaystyle{
		-\al^2 \lxp - 2\al \sqrt{\ml} - \ml
                      \ \ \ (\lxp \geq 1 ) }\\
               \displaystyle{
		-\left(\al + \sqrt{\ml} \right)^2
			\lxp  \ \ \ \ \ \ \ \ \ (\lxp \leq 1 )}
                    \eal  \right. \  .
\ee
These lines are of type 1 ($(\alpha^2-\delta)^2/\alpha^2\gg
4\epsilon/\l$) discussed in section 3.2.  They have one free
parameter, the slope $\al^2$. At spacelike infinity, these
lines are approximately constant Schwarzschild time lines,
$\sigma^0\qe -\ln\alpha$, and for different values of $\al$,
they provide a foliation of spacetime in a way  often
discussed in the literature \cite{lines} in the context of
the black hole puzzle. They always stay outside the event
horizon, and they cross the shock wave at a Kruskal distance
$\delta\qe 2\al \sqrt{M/\l} + \al^2$ from the horizon. After
crossing the shock wave they continue to the strong coupling
region. For an early time Cauchy surface, the parameter
$\al^2$ is arbitrarily large ($\al^2 \rightarrow \infty$
would make the lines approach $\scrim $). As $\al^2$ becomes
smaller, the lines move closer to the event horizon.
Finally, as $\al^2 \rightarrow 0$, the upper segment
asymptotes to $\scrip$ and to the segment of the event
horizon above the shock wave. This is illustrated in Fig. 4.
We are mostly interested in the S-surfaces that catch a
ratio $\r$ of the Hawking radiation emitted by the black
hole, which fixes the value of $\al$. For $r$ not too close
to 1, the S-surfaces are well within the weak coupling
region.

\ifx\answ\bigans
\medskip
\begin{center}
\leavevmode
\epsfxsize 3in
\epsfbox{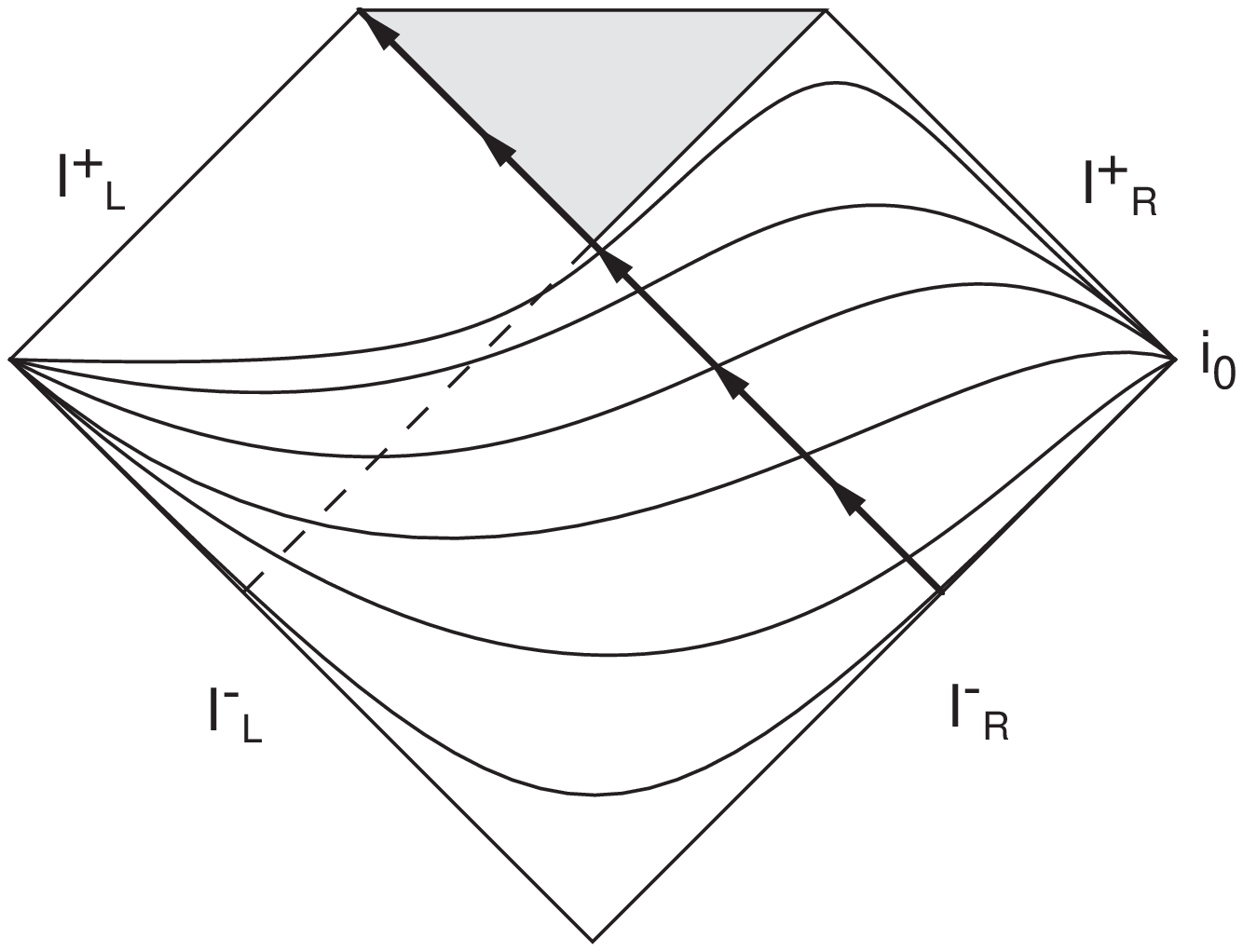}
\end{center}
\vskip 0.5 true cm
\begin{quotation}
\noindent
\small {\bf Figure 4:} Examples of the complete slices of
Sec. 3.3.
\end{quotation}
\medskip
\else
\fi

We want to find the location of the above lines in a black
hole background with a mass $\bar M \qe M + \Delta M $. It is
easy to see that in the new background, the lines
\be
      \lxbm \qe \left\{ \bal
\displaystyle{
		-\alb^2 \lxbp - 2\alb \sqrt{\mbl} - \mbl
                      \ \ \ (\lxbp \geq 1 )} \\
\displaystyle{              -\left(\alb + \sqrt{\mbl} \right)^2
		\lxbp \ \ \ \ \ \ \ \ \ (\lxbp \leq 1 )
                    }\eal  \right. \  .
\ee
also satisfy $d\phi/ds \equiv -\lam$. We only need to
identify the new slope $\alb^2$ in terms of the old one,
and as before, this is given by the boundary conditions at
infinity. Requiring $\lam \bar{\sigma}^+_0 \qe \lam \sigp_0$,
where $\sigma^+$ is the tortoise coordinate defined in
(\ref{coord2}), yields
$$
  \alb \qe
\al + \left(\sqrt{\ml} - \sqrt{\mbl}\right) e^{-\lam \sigp_0}
\ .
$$
If we also want to require $\lam \bar{\sigma}^-_0 \qe \lam
\sigm_0$, we need to do the fixing at infinity, which of
course sets
$$
\alb\qe\al
$$
After fixing the $\sigma^{\pm}$ coordinates at infinity, we
may check that $\bar{\sigma}^{\pm}$ and $\sigma^{\pm}$ do
not differ appreciably as we approach the point $P$ (still
considered to be in the asymptotic region) along an
S-surface. Taking $\al \sim e^{-4rM/\l}$ and $P$ to be at
$\l\xp_{\0}\sim M e^{4\r M/\l}/\l$, $\l\xm_{\0}\sim -M/\l-M
e^{-4\r M/\l}/\l$ as before, we find that at $P$
\bea
   \lam \bar{\sigma}^+_{\0} - \lam \sigp_{\0}
    &\approx & -{\epsilon\over 2M} \left(\frac{\lam}{M}\right)^{1/2}
\\ \nonumber
   \lam \bar{\sigma}^-_{\0} - \lam \sigm_{\0}
    &\approx & -{\epsilon\over 2M}  \ \ ,
\eea
which is a small deviation. We conclude that if we had fixed
the surface at $P$ instead of infinity, all results would be
qualitatively unchanged, as one would expect.

We can now compute the relationship between $\l\xp$ and
$\l\xbp$. As we saw in the previous subsection, the points
in the original line get ``shifted'' by a large amount in
the new line. It is easy to see that
\bea
   \lxbp &\qe & \lxp + \frac{1}{\al} \left(\sqrt{\ml} - \sqrt{\mbl}
	\right) \\
    &\approx & \lxp - \frac{\epsilon}{2\l\al } \sqrt{\frac{
\lam}{M}} \ .
\label{jump}
\eea
For instance, for $\al \sim e^{-4\r M/\l},\ \epsilon/\l \sim
{\l/M}$ the shift is of the order of
$$
   \lxbp - \lxp \sim -\half \left(\frac{\lam}{M} \right)^{3/2}
e^{4\r M/\l} \ ,
$$
which is huge. Even for $\epsilon/\l\sim e^{-M/\l}$, the
shift can be extremely large. As we will see in section 4,
instead of the relations $\lxbp \qe \lxbp (\lxp )$, we will be
interested in the induced relations between the
asymptotically flat coordinates $\lam \bar{\sigma}^{+} $ and
$\lam \sigma^{+}$, and $\lam \bar{y}^-$ and $\lam y^-$. A
huge shift in the Kruskal coordinate close to the shock wave
will correspond to a big shift in the coordinate
$\l\sigma^+$, in which the metric is flat at $\scrim_R$. As
a consequence, the relation between $\lam \bar{\sigma}^+$
and $\lam \sigp $ is nonlinear, as we will discuss in the
next section.

Finally, let us mention an immediate consequence of this
large shift in the $x^+$ direction. The map of an S-surface
from $\m$ to $\mb$ induces a map from a point $Q$ close to
the horizon to a point $\bar Q$ which is shifted a long way
up the horizon in terms of Kruskal coordinates. A similar
map induced by other surfaces through $Q$ which are not
S-surfaces will not shift $\bar Q$ by a large amount. We
therefore see the presence of large quantum fluctuations
near the horizon in the position of $\bar Q$ in the sense
defined above. These large fluctuations are already a
somewhat unexpected result.

\section{The state of matter on $\Sigma$}

We have seen in the previous section in some detail the
large shift that occurs in the $\xp$ direction when we map a
S-surface $\Sigma$ from a black hole spacetime $\m$ to one
with a mass which differs from $\m$ by an extremely small
amount, even compared with the Planck scale. This appears to
be a large effect, capable of seriously impairing the
definition of a unique quantum matter state on $\Sigma$ in a
semiclassical way. There are, however, many large scales in
the black hole problem, and it is premature to draw
conclusions from the appearance of this large shift in the
Kruskal coordinates, without verifying that there is a
corresponding shift in physical (coordinate invariant)
quantities. An absolute measure of the shift is given by the
asymptotic tortoise coordinate $\sigp$ at $\scrim_R$. The
exponential relationship between $x^+$ and $\sigp$ implies
that the shift is of Planck size for an $\xp$ far from the
shock wave ($\xp/\xp_{\0}\sim 1$, where $x_P$ is again as
defined at the end of section 2), and there is no reason to
expect this to give rise to a large effect. However, for
$\xp/\xp_{\0}\ll 1$ (close to the horizon), the shift in $\l
\sigp$ is of order $M/\l$, an extremely large number.  This
implies that the shift is macroscopic in the sense that, for
example, matter falling into the black hole some fixed time
after the shock wave will end up at very different points on
$\Sigma$, depending on whether we work in $\m$ or $\mb$.
Similarly, identical quantum states on $\scrim_R$ should
appear very different on $\Sigma$ in the two cases, meaning
that the matter state on $\Sigma$ is strongly correlated
with the fluctuations in geometry.

In this section, we will attempt to make the notion of
different quantum states of matter on $\Sigma$ more precise,
allowing us to estimate the scale of entanglement between
the matter and spacetime degrees of freedom. In order to do
this, it is necessary to have a criterion to quantify the
difference between two semiclassical matter states living in
different spacetimes $\m$ and $\mb$, that are identical on
$\scrim$ and are then evolved to $\Sigma$. The heuristic
arguments above show that the expectation values of local
operators can be very different for states in $\m$ and $\mb$
that appear identical on $\scrim$ where there is a fixed
coordinate system through which to compare them. Rather than
look at expectation values of operators, we construct an
inner product
$$
\langle \psi_1,\Sigma,\m\vert\psi_2,\Sigma,\mb\rangle
$$
between Schr\"{o}dinger picture matter states on the same
$\Sigma$ through which states on $\m$ and $\mb$ can be
compared. The inner product makes use of a decomposition in
modes defined using the diffeomorphism invariant proper
distance along $\Sigma$, through which the states can be
compared. Details of this construction can be found in
Appendix A.

An important feature of the inner product is that for a
Planck scale fluctuation $\epsilon$ and for states
$|\psi,\m\rangle$ and $|\psi,\mb\rangle$ that are identical
on $\scrim$ it can be checked that
\be
\langle \psi,\Sigma,\m\vert\psi,\Sigma,\mb\rangle\approx 1
\label{cond}
\ee
on any generic surface $\Sigma$ that does not have a large
shift. This is a necessary condition for the consistency of
quantum field theory on a mean curved background with a mass
in the range $(M,M+\epsilon)$: If states on $\m$ and $\mb$
are orthogonal on $\Sigma$, this is an indication that the
approximate Hilbert space structure of the semiclassical
approximation is becoming blurred due to an entanglement
between the matter and gravity degrees of freedom. Using the
inner product, we now show that matter states become
approximately orthogonal on S-surfaces for extremely small
fluctuations $\epsilon/\l\sim e^{-4r M/\l}$ in the mass of a
black hole, dramatically violating condition (\ref{cond}).

In general the states that we wish to compare are most
easily expressed as Heisenberg picture states on $\m$ and
$\mb$, and the prospect of converting these to Schr\"odinger
picture states, and evolving them to $\Sigma$ is rather
daunting. As explained in Appendix A, there is a short cut
to this procedure. For the states we are interested in
(those that start as vacua on $\scrim$) the basic
information needed for the calculation of the inner product
is the relation induced by $\phi(s)$ on $\Sigma$ between the
tortoise coordinates on $\m$ and $\mb$, namely $\sigp \qe
\sigp (\sigbp )$. This allows us to compute the inner
product between the Schr\"{o}dinger picture states by
computing the usual Fock space inner product between two
different Heisenberg picture states, defined with respect to
the modes $e^{-i\om \sigp}$ and $e^{-i\om \sigbp}$. The
latter inner product is given in terms of Bogoliubov
coefficients. It should be stressed that this is just a
short cut, and that the inner product depends crucially on
the surface $\Sigma$, which is seen in the form of the
function $\sigp\qe\sigp(\sigbp)$.

We will study the overlap
\be
  \bra 0~{\rm in}, \Sigma,\m | 0~{\rm in},\Sigma, \mb\ket
\label{e2}
\ee
where $| 0~{\rm in},\Sigma, \m \ket$ is the matter
Schr\"{o}dinger picture state in spacetime ${\cal M}$ on the
hypersurface $\Sigma$ which was in the natural left moving
sector vacuum state on $\scrim_R$. We shall also use this
quantity to estimate the size of $\epsilon\qe (\bar M-M)$ at
which the states begin to differ appreciably. To evaluate
the inner product (\ref{e2}), we first need to find the
induced Bogoliubov transformation

\be
  v_{\om} \qe \int^{\infty}_0 d\om ' \ [\ \al_{\om \om '}
            \bar{v}_{\om '} + \beta_{\om \om '} \bar{v}^*_{\om '} \ ]
\ee
between the in-modes
\bea
    \bar{v}_{\om} &\qe & \frac{1}{\sqrt{2\om}} \
      e^{-i\om \sigbp} \\ \nonumber
      v_{\om} &\qe & \frac{1}{\sqrt{2\om}} \ e^{-i\om \sigp} \ ,
\eea
where $\sigp$ and $\sigbp$ are related by an induced relation
\be
         \sigp \qe \sigp (\sigbp ) \ .
\label{e3}
\ee

Let us derive the relation (\ref{e3}) above, for the example
of Section 3.3. As (\ref{jump}) shows us, the shift $\lam
\xbp - \lam \xp$ can become large and $\lam \xp$ above the
shock wave maps to $\lam \xbp$ further above\footnote{For
simplicity, we will consider only the case $\epsilon<0$ in
this section. The conclusions will not depend on this
assumption.} the shock wave. As $\lxp$ comes closer to the
shock wave and crosses to the other side, the image point
$\lxbp$ can still be located above the shock wave. Only when
$\lxp$ is low enough under the shock wave, does $\lxbp$ also
cross the shock and go below it. Thus, the relation between
the coordinates is split into three regions:
\be
e^{-\phi} \qe \left\{ \bal
              \displaystyle{
			\al \lxp + \sqrt{\ml} \qe \al \lxbp +
                               \sqrt{\mbl} \ \ \ \ \ \ \
                           (\lxp \geq 1) }\\
              \displaystyle{
			 \left(\al + \sqrt{\ml}
			\right) \lxp \qe \al \lxbp + \sqrt{\mbl}
\ \ \ \
                      \left(1 \geq \lxp \geq
\frac{\al + \sqrt{\mbl}}
                                             {\al + \sqrt{\ml}}
\right)} \\
              \displaystyle{
		\left(\al + \sqrt{\ml}
		\right) \lxp \qe
\left(\al + \sqrt{\mbl} \right) \lxbp \ \ \
                   \left(\frac{\al + \sqrt{\mbl}}{\al
+\sqrt{\ml}} \geq
 			\lxp \geq 0\right)}
                            \eal \right. \  \ .
\label{shifts}
\ee
Rewriting (\ref{shifts}) using the asymptotic coordinates,
we then get the relation (\ref{e3}):
\be
    \lam \sigma^+ \qe \left\{ \bal\displaystyle{
 \ln \left[e^{\lam \sigbp} - \frac{1}{\al}\left(\sqrt{\ml} -
\sqrt{\mbl}\right)\right] \ \ \
        \left(\lam \sigbp \geq \lam \sigbp_\q\right) }\\
\displaystyle{
\ln \left[\left(\frac{\al}{\al + \sqrt{\ml}}\right)
\left(e^{\lam \sigbp} +
                                 \frac{1}{\al} \sqrt{\mbl}
\right) \right]
				\ \ \
        \left(\lam \sigbp_\q \geq \lam \sigp \geq 0\right)} \\
	\displaystyle{
  \lam \sigbp + \ln \left[\frac{\al + \sqrt{\mbl}}
                                             {\al + \sqrt{\ml}}
\right] \ \ \
          \ \ \ \ \ \left(0 \geq \lam \sigbp \right)}
      \eal \right. \ \ \ ,
\label{e4}
\ee
where
\be
  \lam \sigbp_\q \equiv \ln \left[ 1 + \frac{1}{\al}
\left(\sqrt{\ml}
                                    -\sqrt{\mbl}\right)
\right] \ .
\ee

This coordinate transformation is illustrated in Fig. 5. As
can be seen, in the first region (which corresponds to both
points being above the shock) the transformation is
logarithmic. On the other hand, in the third region when
both points are below the shock, the transformation is
exactly linear. The form of the transformation for the
interpolating region when the other point is above and the
other point below the shock should not be taken very
seriously, since it depends on the assumptions made on the
distribution of the infalling matter. For a shock wave it
looks like a sharp jump, but if we smear the distribution to
have a width of {\em e.g.} a Planck length, the jump gets
smoothened and the transformation becomes closer to a linear
one.

The Bogoliubov coefficients are now found to be
\bea
  \al_{\om \om '} &\qe & \frac{1}{2\pi } \sqrt{\frac{\om '}{\om}}
                   \ I^+_{\om \om '} \\ \nonumber
  \beta_{\om \om '} &\qe & \frac{1}{2\pi } \sqrt{\frac{\om '}{\om}}
                   \ I^-_{\om \om '} \ \ ,
\eea
where $I^{\pm}_{\om \om '}$ are the integrals
\be
  I^{\pm}_{\om \om '} \equiv \int^{\infty}_{-\infty}
        d\sigbp \  e^{-i\om \sigp (\sigbp ) \pm i\om '
                                \sigbp}  \ \ \ .
\ee
\ifx\answ\bigans
\medskip
\begin{center}
\leavevmode
\epsfxsize 3in
\epsfbox{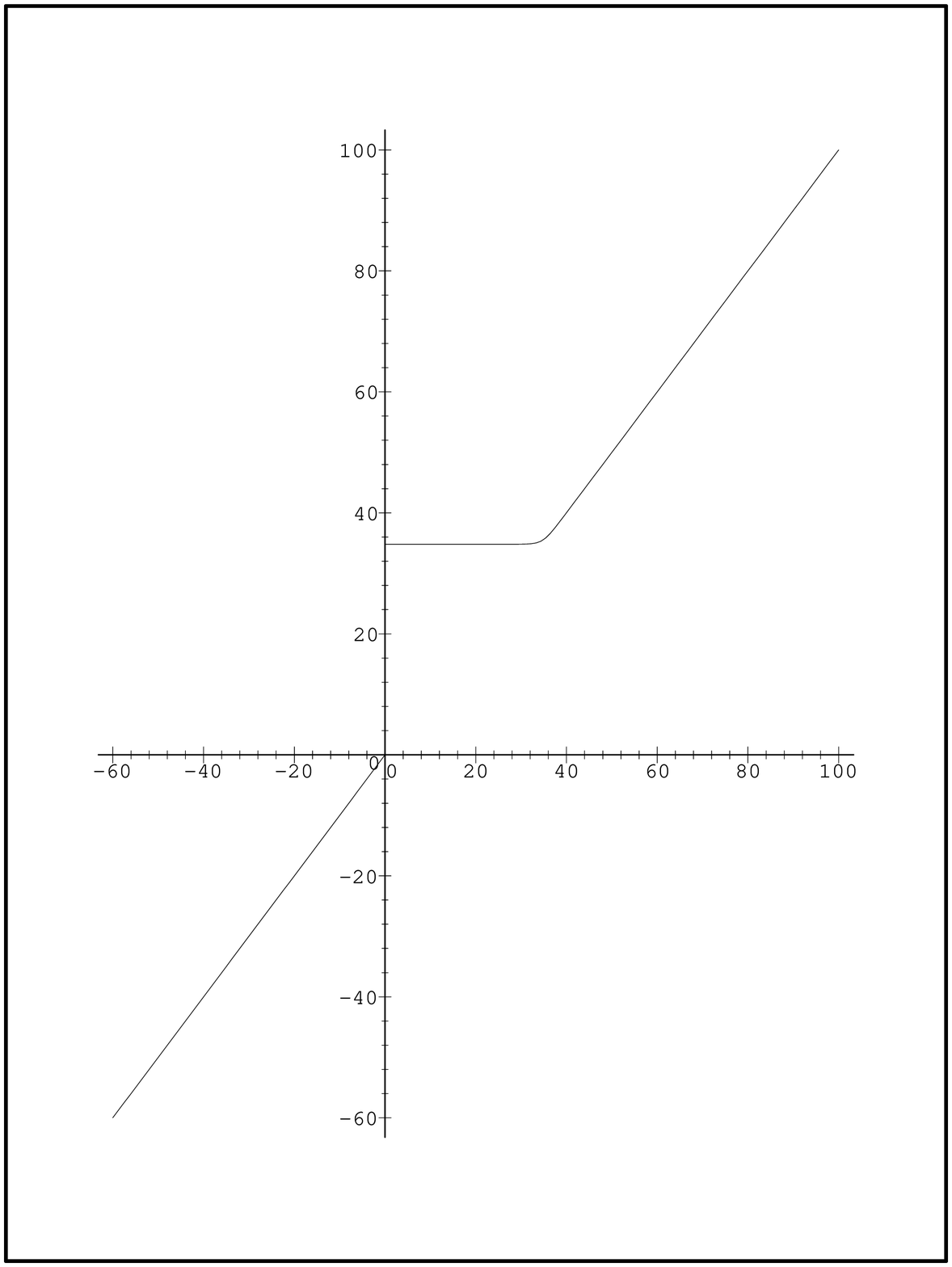}
\end{center}
\vskip 0.5 true cm
\begin{quotation}
\small
\noindent {\bf Figure 5:}
A graph of the function $\l\sigbp(\l\sigp)$. The vertical
axis \qe $\l\sigbp$, the horizontal axis \qe $\l\sigp$. The part
of the graph to the right (left) of the vertical axis
corresponds to both points being above (below) the shock
wave. (The interpolating part is in the region $\l\sigp \in
(-0.0013,0)$ so the plot coincides with a segment of the
vertical axis in the figure). The values $M/\l$ \qe 20,
$\Delta M/\l$ \qe -1/20 and $r$ \qe 1/2 were used in the plot.
\end{quotation}
\medskip
\else
\fi
Substituting the relations (\ref{e4}) above, we get
\bea
\label{integrals}
   I^{\pm}_{\om \om '} &\qe & \int^{\infty}_{\sigbp_\q}
              d\sigbp \ \exp \left\{ -i\frac{\om}{\lam}
        \ln \left[e^{\lam \sigbp} - \frac{1}{\al}\left(
\sqrt{\ml} -
                                   \sqrt{\mbl}\right)\right]
         \pm i\om ' \sigbp \right\} \\ \nonumber
   &+& \int^{\sigbp_\q}_{0}
              d\sigbp \ \exp \left\{ -i\frac{\om}{\lam}
        \ln \left[\frac{\al}{\al +\sqrt{\ml}}
       \left(e^{\lam \sigbp} +\frac{1}{\al} \sqrt{\mbl}
\right)\right]
         \pm i\om ' \sigbp \right\} \\ \nonumber
   &+& \int^0_{-\infty} d\sigbp \ \exp \left\{ -i\om \sigbp
       -i\frac{\om}{\lam} \ln \left[ \frac{\al
+\sqrt{\mbl}}{\al +\sqrt{\ml}}
		\right]
          \pm i\om ' \sigbp \right\} \\ \nonumber
   &\equiv & I^{\pm}_{\om \om '} (1) + I^{\pm}_{\om
\om '} (2)
            +I^{\pm}_{\om \om '} (3) \ \ \ .
\eea

To identify the first integral, introduce first
\be
 \lam \Delta \equiv 1 + \frac{1}{\al} \left(
\sqrt{\ml} - \sqrt{\mbl} \right)\qe e^{\lam
\sigbp_{\q}} \ .
\ee
and change the integration variable from $\lam \sigbp$
to $\lam y$,
\be
    \lam \Delta e^{-\lam y} \equiv e^{\lam \sigbp} \ .
\ee
We now find that
\be
 I^{\pm}_{\om \om '} (1) \qe \lam \Delta^{\pm i \om ' / \lam}
     \int^0_{-\infty} dy \ \exp \left\{ -i\frac{\om}{\lam}
          \ln \left[\lam \Delta (e^{-\lam y} - C) \right]
\mp i\om ' y
\right\} \ \ ,
\ee
where
\be
 C \equiv 1 - \frac{1}{\lam \Delta} \ .
\ee
Now we can recognize the integral to be the same as
discussed in \cite{KM}. This integral can be identified as
an incomplete $\beta$-function. However, it is also possible
to make the standard approximation of replacing the
integrand by its approximate value in the interval $\lam y
\in (-1,0)$ \cite{Haw,KM,gandn}. Note that this interval
corresponds to a region $\lam \sigp \in (0, \ln [(e-1)\lam
\Delta + 1])$. The latter can be large: for $\lam \Delta
\sim e^{M/\l}$ it has size $\sim M/\l$. Indeed, comparing
with (\ref{jump}) we notice that $\lam \Delta  -1$ is equal
to the shift $\lxbp -\lxp$ above the shock, which could
become exponential. So we can use
\be
 I^{\pm}_{\om \om '} (1) \approx \lam \Delta^{\pm i \om ' /
\lam}      \int^0_{-\infty} dy \ \exp \left\{
-i\frac{\om}{\lam}           \ln \left[-\lam \Delta \lam y
\right] \mp i\om ' y \right\} \ \ .
\label{intapprox}
\ee
As discussed in \cite{gandn}, this leads to the approximate
relation
\be
    I^+_{\om \om '}(1) \approx -e^{\pi \om /\lam}
                  \ \left(I^-_{\om \om '}(1)\right)^*
\label{rel1}
\ee
for the integrals.

The logarithm in (\ref{intapprox}) implies that
$I^{\pm}_{\om \om '}(1)$ contributes significantly in the
regime $\om ' -\om \gg 1$. Since the coordinate
transformation (\ref{e4}) was exactly linear in the third
region and we argued that smearing of the incoming matter
distribution smoothens the ``interpolating part'' in the
second region, we can argue that $I^{\pm}_{\om \om '}(2)$
and $I^{\pm}_{\om \om '}(3)$ are negligible in the regime
$\om ' - \om  \gg 1$. Therefore, in this limit $I^{\pm}_{\om
\om '}(1)$ is the significant contribution, and a
consequence of (\ref{rel1}) is that the relationship of the
Bogoliubov coefficients is (approximately) thermal,
\be
   \al_{\om \om '} \approx -e^{\pi \om / \lam}
\ \beta^*_{\om \om '} \ ,
\label{therm}
\ee
with ``temperature'' ${\lam}/{2\pi}$. Let us emphasize that
the ``temperature'' itself is independent of the magnitude
of the fluctuation $\Delta M$. Rather, it is the {\em
validity} of the thermal approximation that is affected: the
larger the fluctuation $\Delta M$ is, the better
approximation (\ref{therm}) is. Also, the region of $\lam
\sigp$ which corresponds to (\ref{therm}) becomes larger.
Consequently, the inner product between $| 0~{\rm
in},\Sigma,\mb \ket$ and $| 0~{\rm in},\Sigma,\m \ket $ can
become appreciably smaller than 1. We refer to this as the
states being ``approximately orthogonal'', we will elaborate
this below.

Let us now calculate the inner product (\ref{e2}). As was
explained before, we have related this inner product to an
inner product between two Heisenberg picture states related
by the above derived Bogoliubov transformation. For the
latter inner product, we can use the general formula given
in \cite{dewitt}. We then find (see Appendices):
\be
  |\bra 0~{\rm in}, \Sigma,\m | 0~{\rm in},\Sigma,\mb \ket|^{2} \qe
      \left(\det (1+\beta\beta^{\dagger})\right)^{-\half}
      \ \ .
\ee
We can now make an estimate of the scale of the fluctuations
for the onset of the approximate orthogonality. As a rough
criterion, let us say that as
\be
    | \bra 0~{\rm in}, \Sigma,\m| 0~{\rm in}, \Sigma,\mb \ket |^2
    \ < \ \frac{1}{\gamma} \ \ ,
\ee
where $\gamma \sim e$, the states become approximately
orthogonal. As is shown in Appendix B, the states become
approximately orthogonal if
\be
  |{\epsilon\over\l} | \qe \left\vert\ml -
{\bar M\over\l }
\right\vert \ > \ \gamma^{48} \ \sqrt{\ml} \ \alpha \ \ ,
\label{criterion}
\ee
where $\alpha $ is the (square root of the) slope. If the
lines do not catch the Hawking radiation, $\al > 1$, then
the fluctuations are not large enough to give arise to the
approximate orthogonality and therefore (\ref{cond}) is
satisfied. On the other hand, if the lines catch the
fraction $\r$ of the Hawking radiation, $\alpha \approx
e^{-4\r M/\l}$ and the fluctuations can easily exceed the
limit. (Recall that $M/\l \gg 1$, so $\al$ is the
significant factor.) Note that the criterion
(\ref{criterion}) has been derived for the example
hypersurfaces of section 3.3. However, the more general
result for any S-surface of the types 1-3 of section 3.2 can
be derived equally easily. In general the right hand side of
(\ref{criterion}) will depend on both the slope $\al^2$ and
the intercept\footnote{Recall that the hypersurfaces of
section 3.3 had $\delta \qe \al^2 + 2\al \sqrt{M/\l}$ so the
rhs of (\ref{criterion}) depends only on $\al$.} $\delta $.
The physics of the result will remain the same as above: for
the S-surfaces the approximate orthogonality begins as the
fluctuations $\Delta M/\l$ satisfy $\ln(\l/\epsilon)\sim
{M/\l}$.

One might ask what happens to the ``in''-vacua at $\scrim_L$
related to the rightmoving modes $e^{-i\om \ym}, e^{-i\om
\ybm}$. We can similarly derive the induced coordinate
transformations between the coordinates $\lam \bar{y}^-$ and
$\lam \ym$. This coordinate relation is virtually linear,
and therefore the $\beta$ Bogoliubov coefficients will be
$\approx 0$ and the vacua will have overlap $\approx 1$.
Thus the effect is not manifest in the rightmoving sector.

\section{Conclusions}

Let us review what we have computed in this paper:

It is widely believed that the semiclassical approximation
to gravity holds at the horizon of a black hole. We have
computed a quantity that is natural in the consideration of
the black hole problem, and that does {\it not} behave
semiclassically at the horizon of the black hole. This
quantity is the quantum state of matter on a hypersurface
which also catches the outgoing Hawking radiation. The
crucial ingredient of our approach was that when we try to
get the semiclassical approximation from the full theory of
quantum gravity, the natural quantity to compare between
different semiclassical spacetimes is the same 1-geometry,
not the hypersurface given by some coordinate relation on
the semiclassical spacetimes. By contrast, in most
calculations done with quantum gravity being a field theory
on a background  two dimensional spacetime, one computes
$n$-point Greens functions, where the `points' are given by
chosen coordinate values in some coordinate system. For
physics in most spacetimes, the answers would not differ
significantly by either method, but in the presence of a
black hole the difference is important.

We computed the quantum state on an entire spacelike
hypersurface which goes up in time to capture the Hawking
radiation, but then comes down steeply to intersect the
infalling matter in the weak coupling region near the
horizon. We found that quantum fluctuations in the
background geometry prevent us from defining an unambiguous
state on this S-surface. Matter states defined on an
S-surface, evolved from a vacuum state at $\scrim$, are
approximately orthogonal for fluctuations  in mass of order
$e^{-M}$ or greater, a number much smaller than the size of
fluctuations expected on general grounds.

One can expand the solution of the Wheeler--DeWitt equation
in a different basis, such that for each term in this basis
the total mass inside the hole is very sharply defined. If
one ignores the Hawking radiation, then one finds that for
such sharply defined mass the infalling matter has a large
position uncertainty and cannot fall into the hole. Thus one
may say that if one wants a good matter state on the
S-slice, then the black hole cannot form. Any attempt to
isolate a classical description for the metric while
examining the quantum state for the matter will be
impossible because the `gravity' and matter modes are highly
entangled. It is interesting that if we try to average over
the `gravity states' involved in the range $M \rightarrow M
+ \Delta M$, we generate entanglement entropy between
`gravity' and matter. This entropy is comparable to the
entanglement entropy of Hawking pairs.

The computations of sections 3 and 4 show that the states on
an S-surface differ appreciably in the region around the
horizon. However, to calculate any local quantity close to
the horizon, we could equally well have computed the state
on spacelike hypersurfaces passing through the horizon
without reaching up to $\scrip$. On these surfaces we would
find an unambiguous state of matter for black holes with
masses differing on the Planck scale. This feature may
signal that an effective theory of black hole evaporation
might not be diffeomorphism invariant in the usual way. It
also indicates that the breakdown in the semiclassical
approximation is relevant only if we try to detect {\em
both} the Hawking radiation and the infalling matter.
Susskind has pointed to a possible complementarity between
the description of matter outside the hole and the
description inside. 't Hooft and Schoutens {\it et. al.}
have expressed this in terms of large commutators between
operators localized at $\scrip$ and those localized close to
the horizon. These notions of complementarity seem to be
compatible with our results. It is interesting that we have
arrived at them with minimal assumptions about the details
of a quantum theory of gravity.

It should be mentioned that although every spacelike
hypersurface that captures the Hawking radiation and the
infalling matter near the horizon gives rise to the effect
we have described, a slice that catches the Hawking
radiation, enters the horizon high up, and catches the
infalling matter deep inside the horizon can be seen to
avoid the large shift. It seems that the quantum state of
matter should be well defined on such a slice. The
significance of this special case is not yet clear to us,
although it is interesting that this type of slice appears
to catch not only the Hawking pairs outside the horizon, but
also their partners behind the horizon.

Our overall conclusion is that one must consider the entire
solution of the quantum gravity problem near a black hole
horizon, in particular one must take the solution to the
Wheeler--DeWitt equation rather than its semiclassical
projection.  We believe that the arguments we have presented
can be applied equally well to black holes in any number of
dimensions.

\section*{Acknowledgements}

We would like to thank R. Brooks, M. Crescimanno,
J. Cohn, A. Dabholkar, S. Das, S. P. de Alwis,
J. Goldstone, A. Guth, J. J. Halliwell,
B. L. Hu, R. Jackiw, P. Kelly, K. Kucha\v{r}, G. Kunstatter,
J. Samuel, A. Sen, C. Stephens, L. Susskind, S. Trivedi,
H. Verlinde, A. Vilenkin, B. Zwiebach and especially
J. G. Demers for discussions.

\appendix

\section*{Appendix A}

In this appendix we shall explain the construction of a
natural inner product relating states of a quantum field
defined on different spacetimes in which the same
hypersurface $\Sigma$ is embedded. Clearly the Schr\"odinger
picture allows us to compute the value of a state on any
hypersurface $\Sigma$ in the form
$$
\Psi[f(x),t_0]
$$
where in the chosen coordinate system $\Sigma$ is the
surface $t\qe t_0$. There is of course a Hamiltonian operator
which is coordinate dependent, and which in the chosen
coordinate system specifies time evolution on constant $t$
hypersurfaces:
$$
H[f(x),\pi_f(x),x,t]\Psi\qe i\dot{\Psi}
$$
We shall assume that the evolution of a state is independent
of the coordinate system used, in the sense that a state on
a Cauchy surface $\Sigma_1$ is taken to define a {\it
unique} state on a later surface $\Sigma_2$. This may not
always be the case \cite{kuchar}, but we will ignore such
problems in our reasoning.

In quantum field theory on a fixed background, there is an
inner product on the space of states on any hypersurface
$\Sigma$. However, in order to use this inner product to
compare states in different spacetimes, it is necessary to
find a natural way of relating two states defined on
$\Sigma$ without reference to coordinates. A natural way of
doing this is to use the proper distance along $\Sigma$ to
define a mode decomposition, and to compare the states with
respect to this decomposition.

Define
\begin{eqnarray}
a(k)&\qe &\int {ds\over \sqrt{4\pi\omega_k}} \;
e^{iks}\left(\omega_k f(x(s)) + {\delta\over\delta f(x(s))}\right)
\nonumber
\\
a^\dagger(k)&\qe &\int {ds\over \sqrt{4\pi\omega_k}}\;
e^{-iks}\left(\omega_k f(x(s))-
{\delta\over\delta f(x(s))}\right)
\label{a2}
\end{eqnarray}
so that $[a(k),a^\dagger(k')]\qe \delta(k-k')$. It is then
straightforward to define the familiar inner product on the
corresponding Fock space. An easy way to picture the Fock
space in this Schr\"odinger picture language is to transform
to a representation $\Psi[a^\dagger(k),t]$, in which $a(k)$
is represented as $\delta/\delta a^\dagger(k)$. The
``vacuum" state, annihilated by all the $a(k)$, is just the
functional $\Psi\qe 1$, and excited states arise from
multiplication by $a^\dagger(k)$. The inner product is
\be
(\Psi_1,\Psi_2)\qe \int {\prod_k da(k) da^\dagger(k)\over 2\pi i}
\left(\Psi_1[a^\dagger(k),t]\right)^*\Psi_2[a^\dagger(k),t]
\exp \left[-\int {dk}\; a(k) a^\dagger(k)\right]
\label{a3}
\ee
where $(a^\dagger(k))^*\qe a(k)$ \cite{fadeevslavnov}.

We could carry out the same construction for any coordinate
$x$ on $\Sigma$, and the inner products would necessarily
agree. However, the operators $a(k)$ and $a^\dagger(k)$
constructed using proper distance are special, in that we
shall say that two states $\Psi_{\m}[f(x),t_1]$ and
$\Psi_{\mb}[f(x),t_2]$ defined on different spacetimes $\m$
and $\mb$ but on a common hypersurface located at $t_\m\qe t_1$
or  $t_{\mb}\qe t_2$, are the same if they are the same Fock
states with respect to this decomposition. If they are not
identical, their overlap is given by the Fock space inner
product (\ref{a3}), and this is what is meant in Section 4
by
$$
\bra\psi_1,\Sigma,\m\vert\psi_2,\Sigma,\mb\ket.
$$

Having defined an inner product between two Schr\"odinger
picture states on different spacetimes, we want to extend it
to Heisenberg picture states on $\m$ and $\mb$, since these
are the kind of states we are used  to working with in
quantum field theory in curved spacetime. The inner product
we have just defined can be used to relate Heisenberg
picture states by transforming each of the states to the
Schr\"odinger picture, evolving them to the common
hypersurface $\Sigma$, and computing the overlap there. It
is useful to have a short-cut to this computation. In order
to achieve this, we first relate a Schr\"odinger picture
state $\Psi[a^\dagger(k),t_0]$ to a Heisenberg picture state
$\Psi[a^\dagger(k)]$, where now the $a(k)$ are associated
with mode functions on $\m$ not $\Sigma$: First choose
coordinates $(x,t)$ on the spacetime such that the metric is
conformally flat, that  $\Sigma$ is a constant time slice,
$t\qe t_0$, and that the conformal factor is unity on $\Sigma$.
Using these coordinates, we can compute the Hamiltonian,
which by virtue of two dimensional conformal invariance is
free
$$
H\qe \int dx (\pi_f^2+(f')^2).
$$
We pick a mode basis defined by
\begin{eqnarray*}
a(k)&\qe &\int {dx\over \sqrt{4\pi\omega_k} }
\; e^{ikx}\left(\omega_k f(x) + {\delta\over\delta f(x)}\right)
\\
a^\dagger(k)&\qe &\int {dx\over \sqrt{4\pi\omega_k}}
\; e^{-ikx}\left(\omega_k f(x)-
{\delta\over\delta f(x)}\right)
\end{eqnarray*}
so that on $\Sigma$ this is precisely the proper distance
mode decomposition. In terms of these modes, the Hamiltonian
is simply given by
$$
H\qe \int dk \omega_k a^\dagger(k) a(k)
$$
so that transforming the operators $a(k)$ and $a^\dagger(k)$
to the Heisenberg picture simply gives
$$
a(k,t)\qe e^{i\omega_k(t-t_0)}a(k),\qquad
a^\dagger(k,t)\qe e^{-i\omega_k(t-t_0)}a^\dagger(k)
$$
and
$$
f(x,t)\qe \int {dk\over\sqrt{4\pi\omega_k}}\left(a(k)
e^{-ik\cdot x}
+a^\dagger(k)e^{ik\cdot x}\right).
$$
Correspondingly the Schr\"odinger picture state
$\Psi[a^\dagger(k),t]$ is identical in form to the
Heisenberg picture state: $\Psi[a^\dagger(k)]\equiv
\Psi[a^\dagger(k),t]\vert_{t\qe t_0}$. We may repeat this
procedure on another spacetime $\mb$, again defining modes
on $\mb$ so that $\Psi_{\mb}[a^\dagger(k)]$ is identical to
the Schr\"odinger picture state on $\Sigma_\mb$. Then, the
inner product (\ref{a3}) serves as an inner product for
Heisenberg picture states $\Psi_\m[a^\dagger(k)]$ and
$\Psi_{\mb}[a^\dagger(k)]$ living on spacetimes $\m$ and
$\mb$.

Now in order to compare a Heisenberg picture state
$\Psi_\m[a^\dagger(k)]$ to any other Heisenberg  picture
state on $\m$, we may make use of standard Bogoliubov
coefficient  techniques. Consider another state defined in
the Heisenberg picture in terms of mode-coefficients related
to modes $v_p(x,t)$ on $\m$. Let us suppose that we
associate operators $b(p)$ and $b^\dagger(p)$ with the modes
$v_p(x,t)$. Then
\begin{eqnarray}
b(p)&\qe & \sum_k \left[\alpha_{kp} a(k) + \beta_{kp}^*
a^\dagger(k)\right]
\nonumber\\
b^\dagger(p)&\qe & \sum_k \left[
\beta_{kp} a(k) + \alpha^*_{kp}a^\dagger(k)\right]
\label{a4}
\end{eqnarray}
where
$$
\alpha_{kp}\qe -i\int dx f_k(x,t)\lra{\partial_t}v^*_p(x,t),
\qquad
\beta_{kp}\qe i\int dx f_k(x,t)\lra{\partial_t}v_p(x,t).
$$
Here $f_k(x,t)\qe e^{-ik\cdot x}/\sqrt{4\pi\omega_k}$ are the
modes defining the $a(k)$.

We may perform a similar calculation on a neighbouring
spacetime $\mb$, also containing $\Sigma$, to relate a set
of modes $\bar{v}_q(\bar{x},\bar{t})$ to the modes
$f_k(\bar{x},\bar{t})$ and similarly to relate the operators
$a(k)$ and $a^\dagger(k)$ to the $\bar{b}(p)$ and
$\bar{b}^\dagger(p)$ as
\begin{eqnarray}
\bar{b}(p)&\qe & \sum_k \left[
	\bar{\alpha}_{kp} a(k) + \bar{\beta}_{kp}^*
	a^\dagger(k)\right]
\nonumber\\
\bar{b}^\dagger(p)&\qe & \sum_k \left[\bar{\beta}_{kp}
	a(k) + \bar{\alpha}^*_{kp}a^\dagger(k)\right]
\label{a5}
\end{eqnarray}
Now we can use the inner product (\ref{a3}) to relate two
states $\Psi_\m[b^\dagger(p)]$ and
$\Psi_{\mb}[\bar{b}^\dagger(p')]$ directly.

More simply, it follows from (\ref{a4}) and (\ref{a5}) that
the $b$'s and $\bar{b}$'s are related by
\begin{eqnarray}
b(p')&\qe & \sum_{p} \left[(\bar{v}_p,v_{p'})\bar{b}(p)
+(\bar{v}^*_p,v_{p'})
\bar{b}^\dagger(p)\right]
\nonumber\\
b^\dagger(p')&\qe & \sum_{p} \left[-(\bar{v}^*_p,v^*_{p'})
\bar{b}^\dagger(p)
-(\bar{v}_p,v^*_{p'})\bar{b}(p)\right]
\label{a6}
\end{eqnarray}
where
\be
(\bar{v}_p,v_{p'})\qe
-i\int_\Sigma dx \bar{v}_p(x,t)\lra{\partial_t}v^*_{p'}(x,t)
\label{a7}
\ee
so that the inner product between states on $\m$ and $\mb$
may be computed using  the standard inner product for states
$\Psi[b^\dagger(p)]$ without going  through the $a(k)$.

In the examples that we consider, the Bogoliubov
coefficients in (\ref{a6}) need not be evaluated on $\Sigma$
as in (\ref{a7}). Suppose for example that we have left
moving mode bases $v_p(\sigp)$ and $\bar v_p(\sigbp)$
defined in terms of tortoise coordinates on $\m$ and $\mb$
respectively. Then both $v_p$ and $\bar v_p$ are functions
of $x^+$ only. We can simply change variables in (\ref{a7})
from $x$ to $\sigma$ (the $t$ differentiation becomes an $x$
differentiation which absorbs the factor $dx/d\sigma$),
yielding
\be
(\bar{v}_p,v_{p'})\qe
-i\int d\sigma
\bar{v}_p(\sigbp(\sigp))\,\lra{\partial_{\sigma^0}}\,v^*_{p'}(\sigp)
\label{a8}
\ee
where $\sigbp$ is given as a function $\sigp$ through the
relations derived by equating points on $\Sigma$ according
to the values of  $\phi$ and $d\phi/ds$, as in Section 3.
The integral (\ref{a8}) looks exactly like the familiar
integral for Bogoliubov coefficients, even though it
involves mode functions on different manifolds. (\ref{a8})
may be evaluated on any Cauchy surface in $\m$ (or $\mb$)
since both the mode functions solve the Klein-Gordon
equation on $\m$ ($\mb$).

\section*{Appendix B}

We present some details of the calculation of the overlap of
the two states on $\Sigma$. We now know the Bogoliubov
transformation between the modes $v_{\om}$ and
$\bar{v}_{\om}$ in the text. Subsequently, the overlap of
the two Schr\"{o}dinger picture states can be found to be
\be
   \bra 0~{\rm in}, \Sigma,\m | 0~{\rm in},\Sigma, \mb \ket
   \qe  (\det (\alpha ))^{-\half} \ \ ,
\ee
where $\al $ is the matrix $(\al_{\om \om '})$ of Bogoliubov
coefficients. The right hand side is the general formula for
the overlap of two vacuum states related to modes connected
by a Bogoliubov transformation \cite{dewitt}. However, it is
more convenient to consider not the overlap but the
probability amplitude
\be
  | \bra 0~{\rm in}, \Sigma,\m | 0~{\rm in},\Sigma, \mb \ket |^2
   \qe  (\det (\alpha \alpha^{\dagger} ))^{-\half} \ \ ,
\label{overlap}
\ee
where the components of the matrix $\al \al^{\dagger}$
are
\be
   (\al \al^{\dagger} )_{\om \om '}
   \qe  \int^{\infty}_0  d\om '' \
     \al_{\om \om ''} \al^*_{\om ' \om ''}  \ \ .
\ee
The evaluation of the determinant of the matrix $\al
\al^{\dagger}$ becomes easier if we move into a wavepacket
basis. Instead of the modes $v_{\om}$ we use
\be
 v_{jn} \equiv a^{-\half} \ \int^{(j+1)a}_{ja} d\om \
  e^{2\pi i \om n/a} \ v_{\om} \ \ .
\ee
These wavepackets are centered at $\sigp \qe  2\pi n/a$, where
$n\qe \ldots , -1,0,1,\ldots$, they have spatial width $\sim
a^{-1}$ and a frequency $\om_j \approx ja$, where
$j\qe 0,1,\ldots$. For more discussion, see
\cite{Haw,gandn,KM}. In the new basis, the Bogoliubov
coefficients become
\bea
  \al_{jn\om '} \qe a^{-\half} \int^{(j+1)a}_{ja} d\om
             \ e^{2\pi i \om n/a} \ \al_{\om \om '} \\ \nonumber
  \beta_{jn\om '} \qe a^{-\half} \int^{(j+1)a}_{ja} d\om
             \ e^{2\pi i \om n/a} \ \beta_{\om \om '}
\eea
with the normalization
\be
  \int^{\infty}_0 d\om '' \ [\al_{jn\om''} \al^*_{j'n'\om ''}
             - \beta_{jn\om ''} \beta^*_{j'n'\om ''} ] \qe
       \delta_{jj'} \delta_{nn'} \ \ .
\label{b1}
\ee
The thermal relation (\ref{therm}) becomes
\be
 \beta^*_{jn\om '} \approx -e^{-\pi \om_j/\lam} \al_{jn\om '} \
\ .
\label{b2}
\ee
Recall that the validity of the thermal approximation
corresponded to the region $\lam \sigp \in (0,\ln
[1+(e-1)\lam \Delta])$ where $\l \Delta - 1$ was the shift
$\lxbp - \lxp$. Let us denote the size of this region as
$\lam L$. Since the separation of the wavepackets is
$\Delta(\lam \sigp ) \qe 2\pi \lam /a$, we can say that
\be
    n_{max} \qe \frac{\lam L}{\Delta(\lam \sigp )}
        \qe \frac{\ln [1+(e-1)\lam \Delta ]}{2\pi \lam / a}
\ee
packets are centered in this region.

Combining (\ref{b1}) and (\ref{b2}), we now see that
\be
    (\al \al^{\dagger})_{jnj'n'} \approx
      \frac{\delta_{jj'} \delta_{nn'}}{1 - e^{-2\pi \om_j /\lam}}
\ee
for $n,n'$ ``inside'' $\lam L$. For the other values of
$n,n'$ (at least one of them being ``outside''),
\be
  (\al \al^{\dagger})_{jnj'n'} \approx \delta_{jj'} \delta_{nn'} \ .
\ee
We are now ready to calculate the overlap (\ref{overlap}).
We get
\bea
\label{b3}
 \ln \left[| \bra 0~{\rm in},\Sigma ,\m | 0~{\rm in},\Sigma ,\mb
\ket |^2
\right]
&\approx & -\half \left\{ \sum^{({\rm inside})}_{n} \sum_j
        \ln \left[\frac{1}{1-e^{-2\pi \om_j/\lam}} \right]\right.\\
       \nonumber
&& + \left.\ln \left[\prod^{({\rm outside})}_n \prod_j \ 1 \right]
\right\}
	\\ \nonumber
&\qe & \half \ n_{max} \ \sum_j \ln \left[1 - e^{-2\pi ja/\lam}
\right] \ .
\eea
In order to estimate the last term, we convert the sum to an
integral\footnote{Notice that one might like to exclude
frequencies corresponding to wavelengths much larger than
the thermal region $\l L$ and impose an infra-red cut off at
$j_{min}a\sim 1/L$. It turns out that for $\infty >\l
L>2\pi$ ($0<j_{min}a < \l /2\pi$) the effect of imposing
this cut off is negligible. Therefore we can just as well
take the integral over the full range.}:
\bea \label{b4}
  \sum_j \ln \left[1-e^{-2\pi ja/\lam}\right] &\rightarrow &
     \int^{\infty}_0 dj \ \ln \left[1-e^{-2\pi ja/\lam}\right] \\
\nonumber
     &\qe & \frac{\lam}{2\pi a} \left(\frac{-\pi^2}{6} \right) \ \ .
\eea
Now, combining (\ref{b3}) and (\ref{b4}), we finally get
a useful formula for the overlap:
\bea
 | \bra 0~{\rm in},\Sigma ,\m | 0~{\rm in},\Sigma ,\mb
\ket |^2 &\approx &
  \exp \left\{ -\half \frac{\lam L}{2\pi \lam/a} \frac{
\lam}{2\pi a}
            \frac{\pi^2}{6} \right\} \\ \nonumber
     &\qe & e^{ -\frac{1}{48} \lam L}  \ \ .
\eea
Now we can estimate when the overlap is $<\gamma^{-1}$ where
$\gamma$ is a number $\sim e$. The overlap becomes equal to
$\gamma^{-1}$ as
\bea
   48\ln \gamma &\qe & \lam L
         \qe \ln \left[1 + (e-1) \lam \Delta\right] \\
\nonumber
          &\approx &
\ln \left[1 + (e-1) \left(1 - \frac{\epsilon}{2\al\l}
\sqrt{\frac{\lam}{M}}
\right)\right] \ ,
\eea
where we used (\ref{jump}) in the last step. Solving for
$\epsilon$, we get
\be
   {\epsilon\over\l}
   \approx -\frac{2}{e-1} \gamma^{48} \ \sqrt{\ml} \
                \al \ .
\label{limit}
\ee
If $|\Delta M/\l|$ is bigger, the states are approximately
orthogonal. Notice that since we used (\ref{jump}) in the
end, (\ref{limit}) is a special result for the hypersurfaces
of section 3.3. However, it is straightforward to generalize
(\ref{limit}) to any S-surface of section 3.2 by using the
relevant shifts as $\l \Delta -1$ and proceeding as above.
In general the right hand side of (\ref{limit}) will then
depend on both $\al$ and the intercept $\delta$.

\end{document}